\documentclass[12pt]{article}
\pdfoutput=1
\usepackage{graphicx}
\usepackage{color}
\usepackage{cancel}
\usepackage{amsmath}
\usepackage{cite}
\usepackage{subfig}

\newcommand{\lsim}{\raisebox{-0.13cm}{~\shortstack{$<$ \\[-0.07cm] $\sim$}}~} 
\newcommand{\gsim}{\raisebox{-0.13cm}{~\shortstack{$>$ \\[-0.07cm] $\sim$}}~} 
\newcommand{\beq}{\begin{eqnarray}} 
\newcommand{\eeq}{\end{eqnarray}} 
\newcommand{\tb}{\tan\beta} 
 
\begin{document}

\vspace*{5mm} 

\begin{center}

\mbox{\large\bf Interpreting the current Higgs excesses at the LHC}  

\vspace*{2mm}

\mbox{\large\bf in the 2HD+a framework}

\vspace*{5mm}

{\sc Giorgio Arcadi}$^{1,2}$ and  {\sc Abdelhak~Djouadi}$^{3}$ 

\vspace{5mm}

{\small 

$^1$ Dipartimento di Scienze Matematiche e Informatiche, Scienze Fisiche e Scienze della Terra, \\
Universita degli Studi di Messina, V. Ferdinando Stagno d'Alcontres 31, I-98166 Messina, Italy.\\[2mm]

$^2$ INFN Sezione di Catania, Via Santa Sofia 64, I-95123 Catania, Italy. \\[2mm]

\mbox{ \hspace*{-6mm} $^3$ Departamento de F\'isica Te\'orica y del Cosmos, Universidad de Granada,
18071 Granada, Spain.} \\[2mm] 
}
\end{center}

\vspace*{1cm}

\begin{abstract}
There are several excesses of events in current LHC data, yet not exceeding the level of significance which would make them to be considered as firm. They point to the possibility  of the presence of a new Higgs particle in the spectrum, in addition to the already observed  125 GeV state. In particular, there are  excesses involving  a diphoton resonance at invariant masses of about 95 GeV, 152 GeV and 650 GeV and an extra scalar might accompany the recent observation of a toponium at a mass of about 350 GeV.  Several interpretations of these excesses have been attempted in extensions of the Standard Model. In this paper, we aim to explain them in the framework of a two Higgs doublet model supplemented by a relatively light pseudoscalar Higgs boson $a$ which would correspond to the putative resonance in most cases. This realistic  2HD+a scenario  is attractive as it is has the virtue to pass all experimental constraints from high-precision experiments and collider searches and, at the same time, to allow for a viable explanation of the dark matter in the universe.  We first update the present constraints on the model, in particular taking into account the latest results on dark matter and Higgs searches, as well the high-precision measurements, including those from Higgs and flavor physics. We then show that the additional Higgs states with the proper mass spectrum and adjusted couplings to fermions, would explain all the LHC excesses (but individually)  while passing the former experimental constraints as well as the theoretical ones.  
\end{abstract}

\newpage

\subsection*{1. Introduction}

The search for new particles beyond those contained in the Standard Model (SM) of particle physics is the main item on the agenda of the two multipurpose experiments at the LHC, ATLAS and CMS. This is particularly the case of additional Higgs bosons which, in scalar extensions of the SM,  would accompany the one with a mass of 125 GeV observed at the LHC in 2012 \cite{Hdiscovery}. This is also the case of possible neutral, massive and stable (at cosmological scales) particle candidates that would form the Dark Matter (DM) in the universe \cite{DM-reviews}. These could well be connected with the extended Higgs sector and would only interact with it, as is the case of the so-called Higgs-portal DM scenarios \cite{H-portal}. 

Unfortunately, no firm signal of the two types of particles has arisen from the substantial amount of data that the LHC experiments have collected in the last decade at center of mass energies of 7, 8 and 13 TeV \cite{HHunting}. In fact, even worse, this large set of data (about 165 fb$^{-1}$ of total integrated luminosity) in combination with high-precision measurements from LHC and other experiments at lower energy  \cite{ParticleDataGroup:2024cfk,Muong-2:2025xyk,Flavor-review} and direct as well as indirect searches of DM states in astroparticle physics experiments \cite{DM-reviews,LZ:2024zvo,McDaniel:2023bju}, are such that many beyond the SM extensions cannot simultaneously allow for the presence of these  states with weak-scale masses. This is particularly the case of attractive and widely discussed scenarios such as the two-Higgs doublet model (2HDM) \cite{2HDM} or its constrained minimal supersymmetric SM version, the  MSSM \cite{MSSM}, where only small and ``unnatural" regions of parameter space can cope with the presence of DM and additional but not too heavy Higgs bosons. 

There are nevertheless two bemols or flats to this unfortunate situation. A first one is that  there are still extensions, albeit slightly more involved than the two mentioned just above, that remain viable. This is for instance the case of the 2HD+a scenario \cite{2HDa-all,2HDa-pot,Goncalves:2016iyg,2HDa-us} in which the SM Higgs sector is extended to include not only an additional complex doublet scalar field, but also a singlet pseudoscalar field $a$. The latter is coupled so that it alleviates the strong constraints on the DM particle candidate, assumed to be a stable isosinglet fermionic state added to the spectrum. A significant portion of the parameter space of the model passes all present theoretical and experimental constraints and can as well accommodate the DM \cite{2HDa-us}. In fact, it can even address the few excesses that are present in LHC data, which represent the second bemol that we mentioned earlier. 

Indeed, there are some modest signals or excesses in the data that LHC has collected in the previous searches, which, in the next campaign with higher luminosity, might well turn into indisputable discoveries. As a matter of fact, there are first several excesses in the diphoton channel at different invariant masses. A first one, at a mass of 95 GeV, was observed sometime ago by CMS  \cite{CMS:2018cyk,CMS:2024yhz} and recently also reported by ATLAS \cite{ATLAS:2024bjr}, with a combined $\approx 3 \sigma$ local significance; see e.g. Ref.~ \cite{Biekotter:2023oen}.  More recently, CMS also observed excesses in the diphoton plus $b\bar b$ channel \cite{CMS:2023boe} (as well as other channels including missing energy as advocated in Ref.~\cite{Crivellin:2021ubm}) corresponding to a putative resonance with  a mass of about 150 GeV. In fact, this search also  shows an excess  that could indicate the presence of a heavier resonance at a mass of 650 GeV. These types of surpluses are rather difficult to interpret in favored beyond the SM extensions such as the 2HDM or the MSSM, the limiting factor being constraints from Higgs and flavor physics. Many other options have also been tried such as 2HDM extensions with scalar singlets, Higgs triplets, Georgi Machacek, extra-dimensional models and extended supersymmetric models such as the NMSSM; see Ref.~\cite{Crivellin:2021ubm,Arcadi:2023smv,Chen:2025vtg,Bhattacharya:2025rfr,LeYaouanc:2025mpk,Excesses-NMSSM,All-papers} for a list of the relevant literature on the subject.     

Furthermore, the LHC collaboration have rather recently observed with great significance an excess of $t \bar t$ threshold events \cite{CMS:2025kzt,CMS:2025dzq,ATLAS:2025kvb}. It  corresponds to the quasi-bound toponium state $\eta_t$ expected in QCD with a mass $m_{\eta_t} \approx 2m_t = 345$ GeV \cite{Toponium}. However, it is possible that a fraction of the signal could be due to an additional pseudoscalar boson $A$ with a mass $M_A \approx 365$ GeV, as also contemplated by CMS itself \cite{CMS:2025kzt}. A straightforward manner to introduce such an extra contribution, besides introducing a generic one \cite{Djouadi:2024lyv} is to assume a 2HDM with a pseudoscalar $A$ of mass of 365 GeV. However, this specific model is very constrained by experimental data and theoretical requirements \cite{Lu:2024twj}. 

In this paper, we will attempt to interpret the three possible excesses at invariant masses of 95 GeV, 152 GeV and 650 GeV in the diphoton channel and part of the one at $\approx 350$ GeV in the $t\bar t$ channel, in terms of the lighter pseudoscalar $a$ state of the 2HD+a scenario. For the parameter configuration that allows to reproduce the various LHC excesses,  mainly the mass and $a$ couplings to heavy fermions, we will investigate if the model passes the constraints from other particle collider searches, high-precision measurements and DM searches. We will particularly take into account the most recent results: the more accurate theoretical determination \cite{Aliberti:2025beg} and experimental measurement at Fermilab \cite{Muong-2:2025xyk} of the anomalous magnetic moment of the muon released very recently, the improvement limits on the DM direct detection rate obtained by the LUX-ZEPLIN (LZ) experiment \cite{LZ:2024zvo} and the updated constraints from heavy flavors \cite{Flavor-review}, in particular from the decays $B \to X_s\gamma$ \cite{bsgamma} and $B_s \to \mu^+\mu^-$ \cite{bsmumu}. We show that, indeed, all these excesses can be explained in the 2HD+a scenario while coping with the aforementioned constraints and that, in the fortunate case where one of the excesses is not a statistical fluctuation, a definite answer and an indisputable signal would emerge at the high-luminosity LHC which is expected to operate in the next decade, or maybe even before   

The rest of the paper is organized as follows. In the next section, we will briefly describe the main aspects of the 2HD+a model and list all the constraints, both theoretical and experimental, to which it is subject. In section 3, we will delineate the parameter space of the  model in which the three diphoton excesses at 95 GeV, 152 GeV and 650 GeV as well as a contribution to the $t\bar t$ signal at about 365 GeV, can be made possible.  In section 4, a short conclusion and a few words of prospective will be given. 

\subsection*{2. The 2HD+a model and the various constraints on it}

\subsubsection*{2.1 The theoretical set-up}

The main interesting feature of the two-Higgs doublet  plus a light pseudoscalar model, 2HD+a,  is that it allows to induce in a gauge invariant way an interaction the singlet pseudoscalar $a$ boson and the SM fermions, $a \bar f \gamma_5 f$, via the mixing of $a$ with the pseudoscalar $A$ state of the 2HDM.  The scalar potential of the model is given by \cite{2HDa-pot} 
\begin{eqnarray} 
V = V(\Phi_1,\Phi_2) + \frac{1}{2} m_{a_0}^2 a_0^2+\frac{\lambda_a}{4}a_0^4
+\left(i\kappa   a_0 \Phi^{\dagger}_1\Phi_2+\mbox{h.c.}\right)+\left(\lambda_{1P}a_0^2 \Phi_1^{\dagger}\Phi_1+\lambda_{2P}a_0^2 \Phi_2^{\dagger}\Phi_2\right),
\end{eqnarray} 
where $V(\Phi_1,\Phi_2)$ is the usual potential of the two Higgs doublet fields \cite{2HDM}, 
\begin{eqnarray}
 V (\Phi_1,\Phi_2) &= m_{11}^2 \Phi_1^\dagger \Phi_1+ m_{22}^2 \Phi_2^\dagger \Phi_2 - m_{12}^2  (\Phi_1^\dagger \Phi_2 + {\rm h.c.} ) +\frac12{\lambda_1} ( \Phi_1^\dagger \Phi_1 )^2 +\frac12{\lambda_2} ( \Phi_2^\dagger \Phi_2 )^2  \nonumber \, \\ &
+\lambda_3 (\Phi_1^\dagger \Phi_1) (\Phi_2^\dagger \Phi_2) +\lambda_4 (\Phi_1^\dagger \Phi_2 ) (\Phi_2^\dagger \Phi_1) +\frac12 {\lambda_5} [\,   (\Phi_1^\dagger \Phi_2 )^2 + {\rm h.c.} \, ] \, .
\label{eq:scalar_potential}
\end{eqnarray}
We have assumed CP-conservation  and a $Z_2$ symmetry to forbid tree-level flavor-changing neutral currents (FCNCs). After symmetry breaking, when the two doublets acquire non-zero vevs $v_1$ and $v_2$ such that $v_1/v_2=\tan\beta$ and $\sqrt{v_1^2+v_2^2}=v\simeq 246~{\rm GeV}$, the scalar sector will consist of two CP--even $h,H$ states, with $h$ conventionally identified with the observed 125 GeV boson, two charged $H^{\pm}$ bosons and two CP--odd bosons which are mixtures of the original singlet and 2HDM pseudoscalar  states  $a_0$ and $A_0$. They are obtained from the field rotation with angle $\theta$ that transform the $(A^0,a^0)$ current eigenstates into the $(A,a)$  physical CP-odd eigenstates
\begin{equation}
\left(
\begin{array}{c} A^0 \\ a^0 \end{array} \right)= \left( \begin{array}{cc}
 \cos\theta & \sin\theta \\ -\sin\theta & \cos\theta \end{array} \right)  \left(
\begin{array}{c} A \\ a \end{array} \right)\, ,  \   \ {\rm with } \ \tan2\theta= \frac{2 \kappa v}{M_{A}^2-M_{a}^2} \, . 
\end{equation} 
In the physical mass basis, this scalar sector can be  fully described by the following set of parameters: the 5 physical Higgs masses $M_h$, $M_H$, $M_{H^{\pm}}$, $M_A$, $M_a$, three couplings of the scalar potential, namely $\lambda_{1P}$, $\lambda_{2P}$ and $\lambda_3$ appearing in  $V(\Phi_1,\Phi_2)$, and the mixing angles entering $\sin\theta$, $\tan\beta$ and $\cos(\beta-\alpha)$, $\alpha$ being the mixing angle among the two CP-even neutral bosons. 

The couplings of the neutral Higgs bosons to the SM fermions which play a crucial role in our context,  are described by the following Lagrangian in the flavor conserving case (the couplings of the $H^\pm$ bosons to isospin $\pm \frac12$ fermions follow that of the unmixed $A$ state): 
\begin{eqnarray}
\mathcal{L}_{\rm Yuk}=\sum_f (m_f/v) [g_{hff} h \bar f f+g_{Hff} H\bar f f- i g_{A^0ff} \cos\theta A  \bar f \gamma_5 f+i g_{A^0ff} \sin\theta a \bar f \gamma_5 a ] \, . 
\end{eqnarray}
To avoid tree-level flavor-changing neutral currents, a structure is generally  assumed for the couplings of the five Higgs bosons which fall into four types of  2HDM configurations \cite{2HDM} which are summarized in Table \ref{table:2hdm_type}. 
These expressions simplify in the so called alignment limit, $\beta-\alpha=\pi/2$, in which the reduced coupling of the $h$ state reduce to their SM values,  $g_{hff}\!=\!1$, while the couplings of the heavier Higgs  states depend only on $\tan\beta$: 
\begin{equation}
    g_{Hff}=\xi_f,\,\,\,\,g_{Aff}=\cos\theta \; \xi_f,\,\,\,\,\,\,g_{aff}=-\sin\theta \; \xi_f , 
\end{equation}
with the parameters $\xi_f$ having four sets of possible assignments, corresponding to four types of 2HDM,  and which are also given in Table \ref{table:2hdm_type}. The perturbativity of the Yukawa couplings constrain the value of the ratio of vevs $\tan\beta$. As in all possible assignments one has $\xi_t = \cot \beta$, one should have $\tan\beta \gsim 1/3$ in order to fulfill this condition for $g_{Att}$ which is enhanced for low $\tan\beta$ values and becomes larger than unity below the value above. In the Type-X (also customarily dubbed lepton-specific) scenario, a perturbative $g_{A \tau\tau} \propto \tan\beta$ imposes $\tan\beta \lsim {\cal O}(100)$. In the Type-II and the  Type-Y  (also called flipped) scenarios,   the $b$-quark Yukawa couplings are enhanced at large $\tan\beta$ leading to the constraint,  $\tan \beta \lsim 50$--60. Note that if one neglects lepton Yukawa couplings, Type I and Type-X scenarios on the one hand, and Type-II and Type-Y ones on the other hands, are equivalent.  

\begin{table}[h!]
\renewcommand{\arraystretch}{1.4}
\begin{center}
\begin{tabular}{|c|c|c|c|c|}
\hline
~~~~~~ &  Type I & Type II & Type X & Type Y \\ \hline \hline 
$g_{htt}$ & $ \frac{\cos \alpha} { \sin \beta} \rightarrow 1$ & $\frac{ \cos \alpha} {\sin \beta} \rightarrow 1$ & $\frac{ \cos \alpha} {\sin\beta} \rightarrow 1$ & $ \frac{ \cos \alpha}{ \sin\beta} \rightarrow 1$ \\ \hline
$g_{hbb}$ & $\frac{\cos \alpha} {\sin \beta} \rightarrow 1$ & $-\frac{ \sin \alpha} {\cos \beta} \rightarrow 1$ & $\frac{\cos \alpha}{ \sin \beta} \rightarrow 1$ & $-\frac{ \sin \alpha}{ \cos \beta} \rightarrow 1$ \\ \hline
$g_{h\tau\tau} $ & $\frac{\cos \alpha} {\sin \beta} \rightarrow 1$ & $-\frac{\sin \alpha} {\cos \beta} \rightarrow 1$ & $- \frac{ \sin \alpha} {\cos \beta} \rightarrow 1$ & $\frac{ \cos \alpha} {\sin \beta} \rightarrow 1$  \\ \hline\hline
$g_{Htt}$ & $\frac{\sin \alpha} {\sin \beta} \rightarrow -\frac{1}{\tan\beta}$ & $\frac{ \sin \alpha} {\sin \beta} \rightarrow -\frac{1}{\tan\beta}$ & $ \frac{\sin \alpha}{\sin \beta} \rightarrow -\frac{1}{\tan\beta}$ & $\frac{ \sin \alpha}{ \sin \beta} \rightarrow -\frac{1}{\tan\beta}$ \\ \hline
$g_{Hbb}$ & $ \frac{ \sin \alpha}{\sin \beta} \rightarrow -\frac{1}{\tan\beta}$ & $\frac{\cos \alpha}{\cos \beta} \rightarrow {\tan\beta}$ & $\frac{\sin \alpha} {\sin \beta} \rightarrow -\frac{1}{\tan\beta}$ & $\frac{ \cos \alpha} {\cos \beta} \rightarrow {\tan\beta}$ \\ \hline
$g_{H\tau\tau}$ & $\frac{ \sin \alpha} {\sin \beta} \rightarrow -\frac{1}{\tan\beta}$ & $\frac{\cos \alpha} {\cos \beta} \rightarrow {\tan\beta}$ & $\frac{ \cos \alpha} {\cos \beta} \rightarrow {\tan\beta}$ & $\frac{\sin \alpha} {\sin \beta} \rightarrow -\frac{1}{\tan\beta}$ \\ \hline\hline
$g_{A^0tt}$ & $\frac{1}{\tan\beta}$ & $\frac{1}{\tan\beta}$ & $\frac{1}{\tan\beta}$ & $\frac{1}{\tan\beta}$ \\ \hline
$g_{A^0bb}$ & $-\frac{1}{\tan\beta}$ & ${\tan\beta}$ & $-\frac{1}{\tan\beta}$ & ${\tan\beta}$ \\ \hline
$g_{A^0\tau\tau}$ & $-\frac{1}{\tan\beta}$ & ${\tan\beta}$ & ${\tan\beta}$ & $-\frac{1}{\tan\beta}$
\\ \hline
\end{tabular}
\vspace*{.1mm}
\caption{Summary of the possible values, avoiding tree-level flavor changing neutral current, of the couplings of the Higgs bosons in the 2HDM model relative to the SM-like Higgs coupling. Their values in the alignment limit $\beta-\alpha=\pi/2$ are also shown.}
\label{table:2hdm_type}
\end{center}
\vspace*{-10mm}
\end{table}
  
There are strong theoretical constraints on the model, in particular conditions on the quartic Higgs couplings in order to have a scalar potential that is bounded from below \cite{Kanemura:2004mg} (similar to the case of a general 2HDM) as well as requirements from perturbative unitarity on the scattering amplitudes of Higgs into gauge boson processes~\cite{Goncalves:2016iyg}. Requirements on the perturbativity of the various Higgs self-couplings,  which also constrain the mass difference between the $A,H,H^\pm $ states to be  not too large,   can be imposed: $\lambda_{\phi_i \phi_j \phi_k} \lsim {\cal O}(4\pi)$. 

All these constraints have been discussed in e.g. Ref.~\cite{H-portal} and will be fully included in our analysis. An important outcome is that one cannot have an arbitrary mass splitting between the $a$ and $A$ bosons when mixing is present: in the limit  $M_A\gg M_a$ and for a maximal mixing  $\sin2\theta \approx 1$, these induce an  upper bound on $M_A$ of about 1.4 TeV which can, however, be  weakened by lowering the value of $\sin2\theta$.
In the parameterization that we adopted, the angle $\theta$ varies in the range $\pm \pi/4$ but  one can restrict, without loss of generality, to positive values of $\sin\theta$. Our chosen range of variation will be the $0.1  \leq \sin\theta \leq 0.7$.\vspace*{-2mm} 

\subsubsection*{2.2 Experimental constraints  on the model}

The 2HD+a scenario is subject to constraints from collider searches of heavy and light extra Higgs bosons and high-precision measurements in the Higgs, gauge boson and heavy-flavor sectors. Let us briefly summarize the relevant constraints which will be treated in the same but updated way as in Ref.~\cite{2HDa-us} where all relevant details can be found. 

There are first \underline{high-precision electroweak as well as SM-like Higgs measurements} \cite{Deltarho}. Concerning the former, we have performed the same type of analysis illustrated in Ref.~\cite{2HDa-us}, based on the computation of the $S,T,U$ loop-parameters. On general grounds, the strongest deviations with respect to the SM occur for the T parameter, which is related to the $\rho$-parameter. In this case,  the new contributions are automatically set to zero if one assumes mass degeneracy for the heavy Higgs states, namely $M_H \simeq M_A \simeq M_{H^{\pm}} \equiv M_\Phi$ \cite{Arcadi:2020gge}. 
On the other hand, LHC measurements of the couplings of the light $h$ state,  do not exhibit relatively large deviations with respect to the SM predictions, less than about $10\%$ at most. As already mentioned, it is possible to automatically impose SM-like couplings for the $h$ state by invoking  alignment. Deviations from the limit of the latter are, however, still allowed, at low values of $\tan\beta$ for the Type-I configuration of the Yukawa couplings and, to a lesser extent, in the Type-X and Type-Y cases as well,  opening the possibility for some interesting LHC signatures. For this reason, we will not  impose strictly the alignment limit in our numerical study. 

Furthermore, there are \underline{constraints from flavor physics}. There is first the constraint from  the anomalous muon magnetic moment or (g--2)$_\mu$ for which the new experimental value measured at Fermilab \cite{Muong-2:2025xyk} and the most recent theoretical prediction in the SM \cite{Aliberti:2025beg}, agree at the $1\sigma$ level, implying that the additional Higgs contributions in the 2HD+a model should be rather small\footnote{For a study of the constraints from $(g-2)_\mu$ on the 2HD+a model, see from example  Ref.\cite{2HDa-us}.}. This will be easily satisfied in our case since the minimal $a$-boson mass that we will assume will be 95 GeV. In turn, in the considered range of masses that we will consider here, the most relevant come from decays of $B$-mesons and, in particular, the  $B_s \to \mu^+ \mu^-$ and $B \rightarrow X_s\gamma$ processes. Concerning the former one, deviations from the SM predictions arise at the one-loop level from diagrams in which all the Higgs bosons of the model are exchanged. To evaluate this contribution in our case, we have combined the SM  one  using the fitting formula provided in Ref.~\cite{Enomoto:2015wbn} with the one-loop new physics contributions,  adapting the result for the $\mathcal{Z}_2$ symmetric 2HDM obtained in Refs.~\cite{Cheng:2015yfu,Enomoto:2015wbn} which is then compared to the  experimental value provided by the Heavy Flavor Averaging Group \cite{HFLAV:2022esi}. In the case of the radiative decay $B \to X_s\gamma$, the $H^\pm$ boson can have large contributions when its couplings to $tb$ pairs are large, namely in the Type-II and Y scenarios for all $\tan\beta$ values, but also Type-I and X at low $\tan\beta$; an updated analysis sets the very strong bound $M_{H^\pm} \gsim 800$ GeV in these cases \cite{bsgamma}. These constraints as well as others, such as those from lepton universality violation $Z$ and $\tau$ lepton decays will be included by adapting to our case those determined for the 2HDM \cite{Abe:2015oca,Chun:2016hzs}. 

Finally, there are \underline{collider and in particular LHC constraints} from searches for additional  Higgs bosons \cite{HHunting}. The strongest one, that applies particularly in the Type-II scenario at relatively high values of $\tan\beta$ comes from  searches of heavy resonances decaying into $\tau$-lepton pairs, $gg \rightarrow a/A/H \to \tau \tau$ where the main contribution is generated by top- and bottom-quark loops. Strong limits from this search have been obtained in the 2HDM and MSSM scenarios, namely $M_H \approx M_A \gsim 1$ TeV for $\tan\beta \gsim 5$. They will be adapted to our model, but they can be significantly weakened by allowing for invisible decays into DM states of the heavier neutral Higgs bosons, a large mixing between the $a$ and $A$ states which suppresses the production of a given pseudoscalar state and, in the case of the CP-even $H$ boson,  the possibility of $H \to aa$ decays.  In turn,  additional and interesting decay modes such as  $H\rightarrow aZ$ and $A\rightarrow ha$ could occur. There are also constraints on the pseudoscalar $a$ from searches of light resonances decaying into muon pairs \cite{CMS:2019buh}. 

At low  $\tan\beta$ values, say $\tan \beta \lsim 3$, the cross section for the dominant $gg\to a,A,H$ production process is mainly generated by top quark loops and can be significant since the top Yukawa coupling is large. For masses above the $t\bar t$ threshold, the Higgs bosons will very dominantly decay into top quark pairs so that the search $pp \to a/A/H \to t\bar t$ becomes extremely constraining. However, the analysis is a bit complicated as one needs to consider the large QCD background from the process $gg \! \to \! t\bar t$ as well as its interference \cite{tt-interference} with the signal as both have the same initial and final states. This renders the interpretation of the searches more problematic as we will discuss. 

Our numerical analysis accounts for the most recent constraints from LHC \cite{HHunting} and is based on computations of the production cross sections via the program SusHi \cite{SUSHI} combined with the package HDECAY\cite{HDECAY} for the evaluation of the decay branching fractions.

\subsubsection*{2.3 The Dark Matter aspects}

We finally discuss the cosmological aspects related to the DM candidate, which is assumed to be a Dirac fermion (the results are exactly the same in the case where it is of Majorana type) which is isosinglet under the SM gauge group. As it is not charged under ${\rm SU(2)_L}$, the DM has no couplings to gauge bosons and because of a ${\cal Z}_2$ discrete symmetry is introduced to make it stable, it couples to  Higgs bosons only in pairs. By virtue of the initial coupling with the $a_0$ boson, and following symmetry breaking, the DM will interact with the two pseudoscalar Higgs bosons with the following Lagrangian 
\begin{eqnarray}
\mathcal{L}_{\rm DM}=g_\chi \left(a\, \cos\theta +A\, \sin\theta \right) \bar \chi i \gamma_5 \chi \, . 
\end{eqnarray}
The DM phenomenology has been extensively discussed, e.g. in 
Refs.~\cite{DM-reviews,2HDa-us}. Here, we will simply list the main constraints referring for detailed discussions to the references above. First of all, we have the cosmological relic density constraint. We assume the conventional freeze-out paradigm in which the experimentally favored value measured by the Planck collaboration \cite{Planck:2018vyg},  $\Omega_{\chi}h^2 = 0.12 \pm 0.0012$, is achieved if the DM thermally averaged pair annihilation cross section is in the appropriate range. The most relevant annihilation processes are into SM fermion pairs through pseudoscalar Higgs boson exchange and, if kinematically accessible, also annihilation into $aa, Zh$ as well as  $ha$ final states
\begin{equation}
\chi \chi \to a^*, A^* \to \tau^+ \tau^-,  \ b\bar b ,  \  t \bar t , \ \ 
 {\rm and}  \ \ \chi \chi \to a^*, A^* \to   h a, Zh \ ;  \ \chi \chi \to aa \, .  
\end{equation}
For what the computation of the relic density is concerned, we have used the numerical package micrOMEGAs \cite{MicroMegas}, ensuring a fast and precise computation including also the subdominant annihilation processes into the heavy 2HDM bosons. 

Moving to DM direct detection, we remark that since there are no coupling between a DM pair and the CP-even neutral Higgs bosons, spin-independent interactions are forbidden at tree-level but arise at one-loop from the interactions between the DM and the pseudoscalar boson and the trilinear couplings between CP-even and CP-odd bosons.  The elastic-scattering cross section of the DM over nucleons is calculated as indicated in the relatively recent analysis of Ref.~\cite{DD-xsection}. It will be compared with the very stringent experimental constraint released recently by the LUX-ZEPLIN (or LZ) experiment \cite{LZ:2024zvo}. 

Finally, as the DM annihilation rate into SM fermions   is $s$-wave dominated, the 2HD+a model is also sensitive to constraints from DM indirect detection. We will thus include in our analysis the limits from searches of $\gamma$-rays given in Ref.~\cite{McDaniel:2023bju}. Such limits can effectively constrain the thermal freeze-out paradigm for DM masses up to ${\cal O}(100-200)$ GeV\footnote{A very recent analysis \cite{Manconi:2025ogr} found even more stringent constraint from signals from the Galactic Center.}.\vspace*{-3mm}


\subsection*{3. Signals at the LHC for the various resonances}

\subsubsection*{3.1 The 95 GeV diphoton resonance}

The two LHC multi-purpose experiments have observed a mild excess in the diphoton spectrum at a mass of $M_\Phi=95.4$ GeV. CMS finds a combined excess of 2.9$\sigma$ at the local level \cite{CMS:2018cyk,CMS:2024yhz} when the 8 TeV and the 13 TeV data in Run II are added, while ATLAS observes a local excess of only 1.7$\sigma$ in its Run II data \cite{ATLAS:2024bjr}. The two results when roughly combined,  lead to a total excess  of  3.1$\sigma$  \cite{Biekotter:2023oen} which is quite encouraging, yet not enough to claim a strong evidence.  

More precisely, the diphoton signal strength   is found to be $\mu_{\gamma\gamma}=0.24^{+0.09}_{-0.08}$ which, assuming a SM-Higgs cross section in gg-fusion of  {48.5}\;pb at $\sqrt s=13$ TeV and a two-photon branching fraction of ${2.27 \times 10^{-3}}$ \cite{LHC-H-WG}, would lead to the cross section times branching ratio for an $a$ boson decaying into two photons of 
\begin{equation} 
\sigma (gg \to a) \times {\rm BR} ( a \to \gamma\gamma) \simeq {26.5^{+9.9}_{-8.8}~{\rm fb}} @ \sqrt s=13~{\rm TeV}\ \ \ {\rm for} \ M_a=95~{\rm GeV}.  
\label{mu-95} 
\end{equation} 
Here, we will simplify the discussion\footnote{CMS observes also an excess in the $\tau\tau$ channel at a mass of about 100 GeV \cite{CMS:2022goy}, which is compatible with a 95 GeV  resonance at about 2.5$\sigma$; but the channel has not be discussed by ATLAS. There is also the longstanding LEP excess at about the same mass in $e^-e^+ \to Z b\bar b$ which is conflicting with the previous one; see the discussion in Ref. \cite{Chen:2025vtg}. We will therefore ignore both channels in our analysis.}  and assume that only the $a$ state is kinematically accessible and would couple only to fermions. It would be then produced mainly in the $gg$-fusion mechanism through top and bottom quark loops, $gg \to a$, and will decay into two photons through the same process. Following the procedure outlined in Ref.~\cite{Arcadi:2023smv}, we have determined the signal strength of the pseudoscalar 95 GeV resonance in $\gamma \gamma$ as:
\begin{align}
\label{eq:mugamgam}
    &  \mu_{\gamma \gamma} = R_{gg} R_{\gamma\gamma} \times \sigma (gg \to \phi)|_{\rm SM} / \sigma (pp \to \phi)|_{\rm SM} \, , 
\end{align}
where the correction factor for the difference between the total and approximate SM Higgs cross section with $gg$-fusion only, has been included.  In terms of the partial decay width into the $gg$ and $\gamma\gamma$ modes, one has
\begin{align}
    & R_{gg} =\frac{\Gamma(\phi\rightarrow gg)}{\Gamma(\phi\rightarrow gg)|_{\rm SM}}= \frac{|g_{A uu} A^P_{1/2} (\tau_t)+g_{A d d} A_{1/2}^P (\tau)|^2}{| A_{1/2}^S (\tau_t)+ A_{1/2}^S (\tau_b)|^2}  \times  \frac{1+\delta_P^{gg}}{1+\delta_S^{gg}},\nonumber\\
    & R_{\gamma\gamma} =\frac{\Gamma(\phi\rightarrow \gamma \gamma)}{\Gamma(\phi\rightarrow \gamma \gamma)|_{\rm SM}}= \frac{|\frac{8}{3}g_{A u u} A_{1/2}^P (\tau_t)+\frac{2}{3}g_{Add} A_{1/2}^P (\tau_b)+2 g_{A l l} A_{1/2}^P (\tau_\tau)|^2}{| \frac{8}{3}A_{1/2}^S (\tau_t)+ \frac{2}{3}A_{1/2}^S (\tau_b)+2A_{1/2}^S (\tau_\tau)-A_1 (\tau_W)|^2} \times \frac{1+\delta_P^{\gamma\gamma}}{1+\delta_S^{\gamma\gamma}}.
\end{align}
The  standard scalar and pseudoscalar form factors $A_{1/2}^{S,P}$ for the fermionic loop contributions and $A_{1}$ for the spin-one $W$-loop  contributions, with arguments $\tau_X= M_\phi^2/4m_X^2$ can be found in Ref.~\cite{Anatomy1}. The correction factors $\delta^{\gamma}_{S,P}$ that take care of the electroweak corrections, which can become sometimes large,  are instead defined in Ref. \cite{Choi:2021nql}.

Adapting the numerical programs SusHi \cite{SUSHI} for the production rate and HDECAY  \cite{HDECAY} for the  branching ratios, to our 2HD+a scenario, we have delineated the parameter space in which the rate of eq.~(\ref{mu-95}) can be obtained. In fact, the situation is simple when the  alignment limit and a heavy 2HDM spectrum is assumed: the only relevant parameters, when the $a$ mass is fixed to $M_a=95$ GeV, are $\tan\beta$ and the mixing parameter $\sin\theta$. This is particularly true if the DM particle is heavier than about $\frac12 M_a \approx 48$ GeV: it will not alter $a$ phenomenology and will be still compatible with DM  constraints. As will be evidenced in the following, DM constraints will favor relatively heavy candidates. Consequently, the LHC phenomenology which will be discussed here is affected to a negligible extent by the presence of the DM state.   

\begin{figure}[!h]
\vspace*{-3mm}
    \centering
    \subfloat{\includegraphics[width=0.5\linewidth]{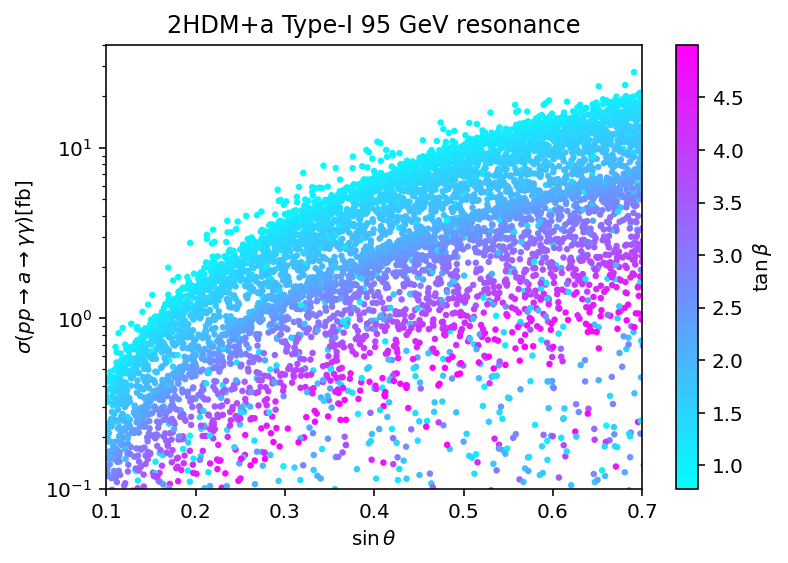}}
    \subfloat{\includegraphics[width=0.5\linewidth]{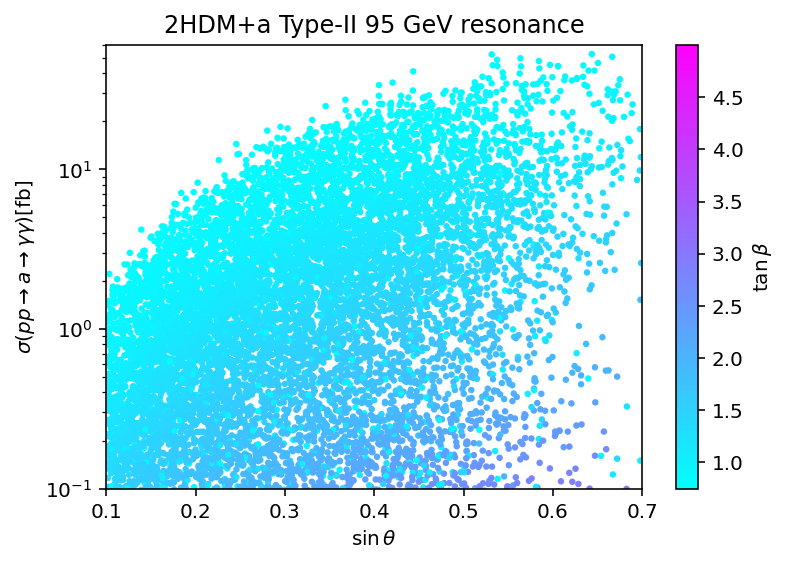}}\\[-3mm]
    \subfloat{\includegraphics[width=0.5\linewidth]{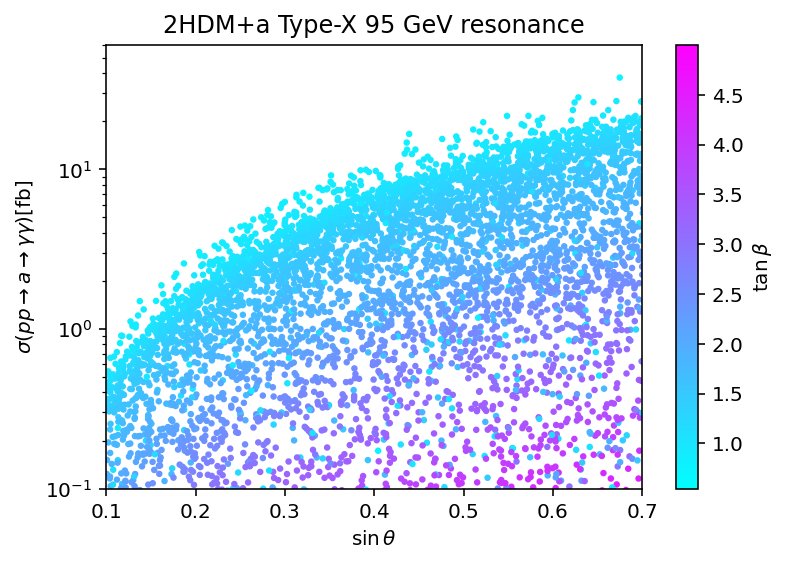}}
    \subfloat{\includegraphics[width=0.5\linewidth]{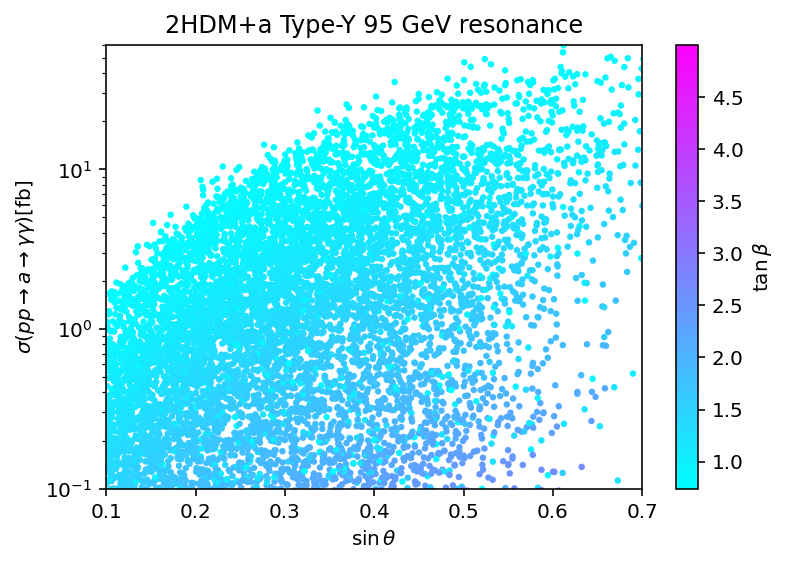}}
 \vspace*{-3mm}
    \caption{\small Production cross section of a 95 GeV pseudoscalar resonance (interpreted as the lightest $a$ state in the 2HD+a model) decaying into $\gamma \gamma$ as a function of $\sin\theta$. The scatter plot has been obtained by performing a parameter scan when accounting for theoretical constraints  and experimental constraints from precision Higgs and electroweak measurements as well as flavor physics. The four different panels correspond, as reported on their top, to the Type-I, Type-II, Type-X and Type-Y. The color pattern of the model points follow corresponding values of $\tan\beta$.}
    \label{fig:95gamgam}
\vspace*{-3mm}
\end{figure}

We have performed a scan of the 2HD+a parameter space, in the four models types for the Yukawa coupling configurations,  varying the relevant parameters in the range. 
\begin{eqnarray}
\label{eq:scan95}
    M_{H,A,H^{\pm}} \gsim [0.1,1.5\,\mbox{TeV}], \   \left \vert \cos\left(\beta-\alpha\right)\right \vert \leq 0.2  , \ \ \sin\theta \in [0.01,0.707],\ \  \tan\beta \in [0.3,10] \, .
\end{eqnarray}
The results are shown in Fig.~\ref{fig:95gamgam} where we plot the production rate times decay branching fraction, $\sigma (pp \to a) \times {\rm BR}(a \to  \gamma\gamma)$ [in fb], as a function of the sine of the mixing angle $\sin\theta$ (for which we show only the range $0.1 \leq \sin\theta \leq 0.7$) for the mass value $M_a=95$ GeV, in the four model types; the corresponding values of $\tan\beta$ are also indicated. 

\begin{figure}[!h]
\vspace*{-5mm}
    \centering
    \subfloat{\includegraphics[width=0.48\linewidth]{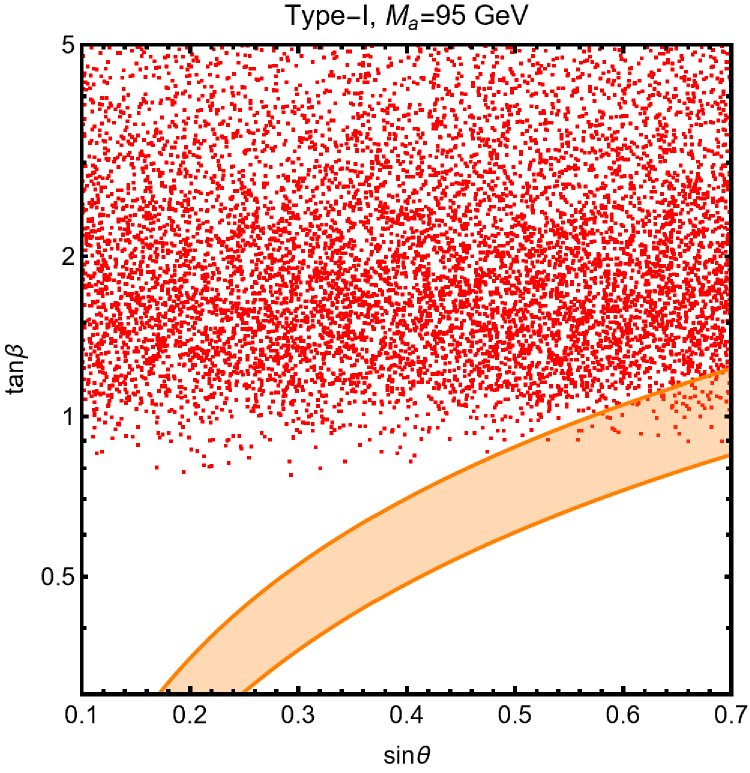}}
    \subfloat{\includegraphics[width=0.48\linewidth]{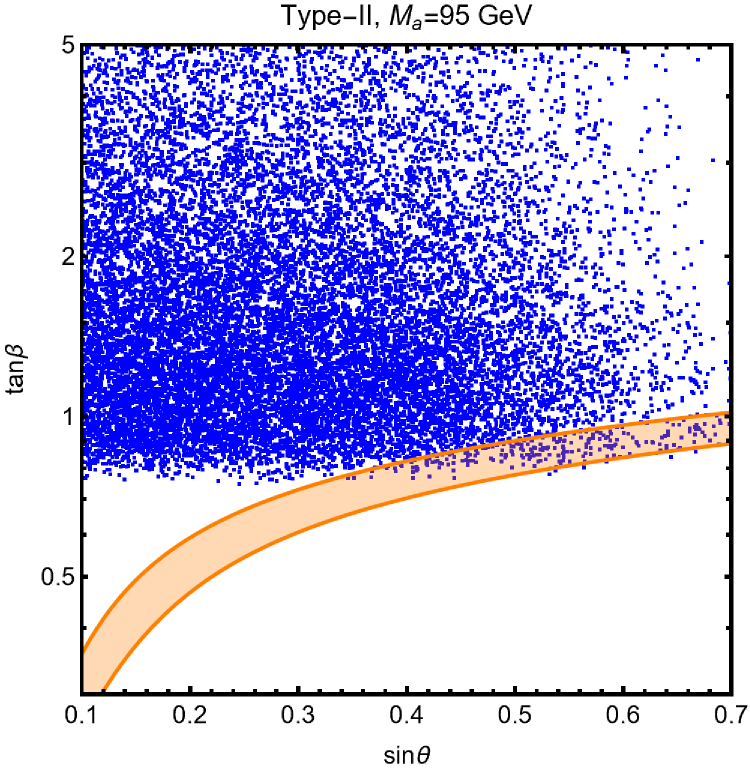}}\\[-3mm]
    \subfloat{\includegraphics[width=0.48\linewidth]{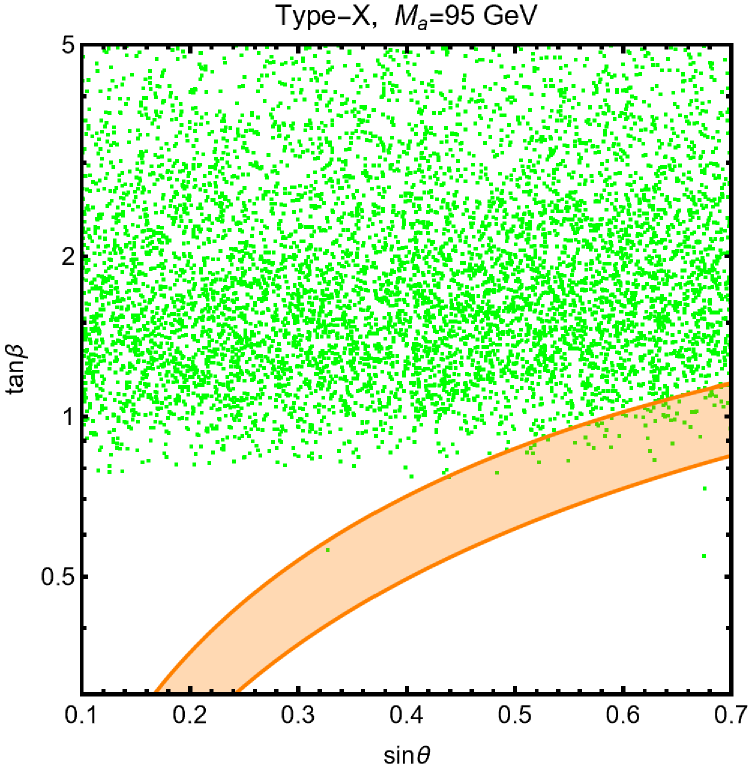}}
    \subfloat{\includegraphics[width=0.48\linewidth]{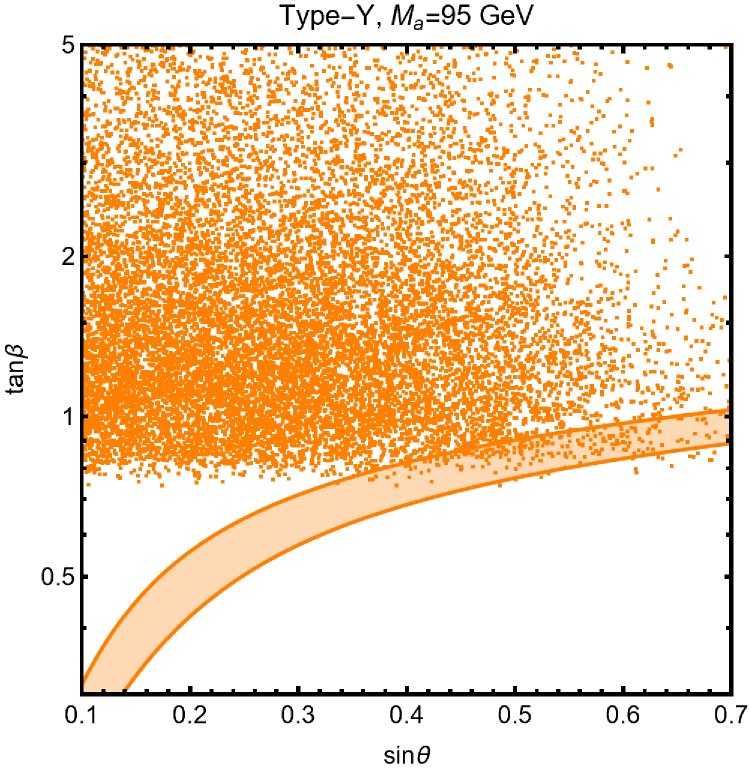}}
\vspace*{-3mm}
\caption{\small Model points in the $[\sin\theta,\tan\beta]$ plane obtained from the parameter scan of eq.~(\ref{eq:scan95}) complying with the general constraints and having $M_a=95\,\mbox{GeV}$. The orange band corresponds to the region compatible with the experimental hint of a 95 GeV resonance decaying into $\gamma \gamma$.}
\label{fig:plot95}
\vspace*{-3mm}
\end{figure}

The possibility of reproducing the experimental excess is illustrated more explicitly in Fig. \ref{fig:plot95} which shows the parameter assignments in the $[\sin\theta,\tan\beta]$ parameter plane which complies  with all theoretical and experimental constraints, including those from flavor physics. Again, the four different panels correspond to the four FCNC preserving Yukawa configurations. We also show as an orange band, the area corresponding to the $[\sin\theta,\tan\beta]$ values fulfilling the condition  $0.16 \leq \mu_{\gamma \gamma} \leq 0.32$ for  the combined CMS and ATLAS diphoton searches at a mass of 95 GeV \cite{Biekotter:2023oen}. 

As can  be seen from Fig.~\ref{fig:95gamgam}, a few points for large  mixing $\sin \theta \gsim 0.5$ and values of $\tan\beta$ slightly below unity, meet all requirements in Type-I and X scenarios, while a much larger range of parameters could fit the signal in models of Type-II and Y with enhanced Yukawa couplings at $\tan\beta$ values above unity. 
The outcome of Fig.~\ref{fig:plot95} can be understood as follows. For what concerns the distributions of points, the near absence of points at $\tan\beta \lsim 1$ is  mostly due to the constraint from $B_s \rightarrow \mu^+ \mu^-$ (see Ref.~\cite{Arcadi:2023smv} for an explicit illustration). The reduced density of model points at high $\sin\theta$ in the Type-II and Type-Y cases comes from the lower bound $M_{H^\pm} \geq 800\,\mbox{GeV}$ from $b \rightarrow s \gamma$. This bounds translates into an analogous one on the masses of the $H,A$ bosons as a result of the $\rho$-parameter constraint. 
The latter renders more effective the unitarity bound on $M_A$-$M_a$ splitting. 

As for the shape of the orange contours, they can be understood from the analytical expressions provided in Ref.~\cite{Arcadi:2023smv} which we will briefly summarize. First, one should recall that both the $agg$ and $a\gamma \gamma$ vertices are proportional to $\sin\theta$ and thus, $\mu_{\gamma \gamma}$ is maximized for $\sin\theta=1/\sqrt{2}$, for fixed values of the other parameters. In the case of the Type-I scenario one would have that $R_{\gamma \gamma} \simeq 0.21$, independently of $\tan\beta$, in contrast to the case of $R_{gg}$. One would then obtain for the $\gamma\gamma$ signal strength:
\begin{equation}
    \mu_{\gamma \gamma} \simeq 0.5\; {\sin^2 \theta}/{\tan^2 \beta} \, ,
\end{equation}
and, for its the central value to be  compatible with the reported excess, one needs $\tan\beta\simeq 1$ for $\sin\theta=1/\sqrt{2}$. In the Type-X model, $R_{\gamma \gamma}$ will in turn depend on $\tan\beta$  so that, again for $\sin\theta=1/\sqrt{2}$. one has
\begin{equation}
    \mu_{\gamma \gamma}\simeq {6.75 \cot^2 \beta}/{(12.25+\tan^4 \beta )} \, , 
\end{equation}
also complying with the experimental value for $\tan\beta \simeq 1$. In the case of the Type-II and Type-Y models, we have instead a more complicated dependence on $\tan\beta$ giving:
\begin{equation}
    R_{gg}\simeq {(1.27-0.14 \tan^2 \beta+0.01 \tan^4\beta)}/{\tan^2\beta} \, , 
\end{equation}
and
\begin{equation}
R_{\gamma \gamma}= \left \{
    \begin{array}{cc}
     {0.25}/{(0.24+\tan^4 \beta)}    &  \mbox{for~Type-II} \\
     {0.28}/{(0.37+\tan^4 \beta)}    &  \mbox{for~Type-Y}
    \end{array}
    \right . \, . 
\end{equation}

\subsubsection*{3.2 The 152 GeV resonance}

The excess corresponding to a 152 GeV resonance, accompanied by other heavy states has been discussed mainly in  Refs.~\cite{Crivellin:2021ubm,Bhattacharya:2025rfr}. Here, the combination of different final state events, involving rather complicated topologies with two photons, $Z$-photon, multiple leptons, moderate missing energy as well as bottom quark jets has been studied, leading to an excess with 5.4$\sigma$ global significance according to the authors. The evidence of this 152 GeV resonance is not very well established and requires further independent studies, in particular by the experimental collaborations,  to be confirmed. It is nevertheless legitimate to ask whether the 2HD+a scenario can accommodate an hypothetical signal.

The same references discussed above, \cite{Crivellin:2021ubm,Bhattacharya:2025rfr}, propose an interpretation of the excess in terms of a simplified model in which there is first a heavy scalar boson (dubbed $H$) which is produced in gluon-fusion through a quark loop which and decays into a lighter scalar boson ($S$). The latter then decays into $\gamma\gamma$ or $Z\gamma$ final states and another scalar $S'$, which can be the same $S$, the SM-like $h$ boson or another new scalar stare $S'$. This $S'$ will give the additional topologies with $b$-jets, $E_{T}^{\rm mis}$ and/or multi-leptons (from decays into $W,Z$ bosons).  Such a scenario can occur in a triplet model as discussed in the original reference \cite{Bhattacharya:2025rfr} but it can also occur in an extended Georgi-Machacek model  \cite{Chen:2025vtg} or in other models with extended Higgs sectors like the NMSSM \cite{Excesses-NMSSM}.     

 In the setup that we are scrutinizing here, on might consider the initial production of a heavy neutral CP-odd $A$ state\footnote{An initial charged Higgs produced in association with a top quark and decaying into a light $a$, $gb \to H^-t \to a Wt$ can also give similar topologies but we will ignore this option in our  preliminary investigation.}, with a mass of around 300 GeV, decaying into the SM-like Higgs and  either the light CP-odd singlet like state $a$ or  the $Z$ boson.  The $A\to Zh$ final state is expected to have a more suppressed cross section as the $AZh$ coupling would require a deviation from the alignment limit, which is constrained to be tiny by the Higgs signal strengths. Alternatively one could interpret the 300 GeV state as the heavy CP-even state $H$ decaying into the $aZ$ final. In summary, we are considering the following signals
\begin{eqnarray} 
&& gg \to A \to a h \to (\gamma \gamma) + (bb, WW \to \ell \nu \ell \nu, ZZ \to \ell\ell/ \nu\nu) ,  \nonumber \\
&& gg \to ~~H~ \to a Z \to (\gamma \gamma) + (bb, qq, \ell\ell, \nu\nu) .
\end{eqnarray}

It is clearly beyond the scope of this paper to make a full simulation of all the possible options and the different final state topologies considered in Refs.~\cite{Crivellin:2021ubm,Bhattacharya:2025rfr}. What we will do, instead, is to perform a scan of the parameter space, similar to the one we made before,  and show that the cross sections times branching ratios for the most important topology vary in a sufficiently large domain so that one can easily fit any signal. The main requirement,  besides making sure that all the topologies in which there are excesses at the LHC are possible to generate, would then  simply be to show that the relevant Higgs spectrum is still allowed by theoretical requirements and experimental data and that it provides the correct amount of DM without being excluded by astroparticle experiments. 

If the production times decay rates fall into the wide range that one expects, the exact numerical values of the final state rates can, in principle, be reproduced by the model since we have enough free parameters. Indeed, except for the Higgs boson masses (and the mixing angle $\alpha$ in the alignment limit) which are fixed, we have the set $[\tan\beta, \sin\theta]$ that we will vary in the same range as previously, and the trilinear Higgs couplings, in particular $\lambda_{Hah}$ and $\lambda_{Aah}$ which are almost free (except for the requirement that they should remain perturbative). Some additional missing energy can also be generated by allowing some decays into the invisible DM particle, such as $a \to \chi \chi$ but also $A \to \chi \chi$ in the cascades, whose mass and coupling to $a/A$ can be appropriately adjusted.   

Hence, the only strict requirement would be to reproduce the order of magnitude of the observed signal. In Ref.~\cite{Crivellin:2021ubm}, it is indicated that for the inclusive case when the channel $a \to \gamma\gamma+X$ is considered, with $X$ being any other final leptonic (including neutrinos)  or hadronic (including bottom) states but when only gluon fusion is considered for the primary process and all other channels including vector-boson and $b\bar b$ fusion are ignored, the target value for the production rate of the $\gamma\gamma$ resonance should be 
\begin{equation}
\sigma ( pp \to a +X) \times {\rm BR} (a \to \gamma\gamma) = (6.6 \pm 3.2) ~{\rm fb}~@ \sqrt s=13~{\rm TeV}\  \ {\rm for} \ M_a=152~{\rm GeV}  . \label{mu-152} 
\end{equation}
The other final states, $\gamma \gamma+E_T^{\rm mis}$, $\gamma \gamma+V$, $\gamma \gamma + b\!\!-\!\!{\rm jets}$ as well as $\gamma Z + V\to \ell \ell, \nu \nu $ and $b\bar b + E_T^{\rm mis}$ account for a combined cross section of $\sigma(S+X) \simeq 3.4 \pm 2.2$ fb according to Ref.~\cite{Crivellin:2021ubm} (see their Table 2) can be certainly accounted for by the production of the cascade decays of the 2HDM states. So it will be sufficient to obtain the correct rates for the $\gamma\gamma$ final state, eq.~(\ref{mu-152}).  We therefore make a scan of the parameter space and delineate the areas in which this rate is reproduced with an order of magnitude margin. 

Note, finally, that there is a much simpler way to reproduce this particular final state:  it is to assume  only the presence of the light $a$ boson (the other Higgses are to heavy to contribute) and obtain the excess (as well as the $Z\gamma$ excess) simply by producing $a$ in gg-fusion, $gg \to a \to \gamma\gamma, Z\gamma$. To recover some of the additional topologies which have a much smaller cross sections, one could move slightly away for the alignment limit (which is valid only at the 10\% level), assume a small but non-zero $haZ$ coupling  and consider the associated production process  $q\bar q \! \to \! Z^* \!  \to \! ha$. While the $a$ will still decay into $\gamma\gamma$, the SM-like $h$ state would lead  to the  additional $b$-quarks, multi-leptons and missing energy. 

Let us now come to the numerical analysis for which we will simply focus on the Type-I scenario. The reason is twofold. First of all, it is not possible to accommodate a relatively light spectrum, namely masses of a few hundred GeV, for the heavier 2HDM or 2HD+a bosons in the Type-II and Type-Y configurations because the combination of the theoretical and EWPT constraints with the very stringent lower bound on the mass of the charged Higgs boson from $b\rightarrow s \gamma$ transitions favors $M_{H^\pm} \approx M_H \approx M_A \gsim 800$ GeV \cite{bsgamma}. In addition, the Type-I configuration is the only one that allows for a non negligible deviation from the alignment limit \cite{2HDa-us}, ensuring a richer LHC phenomenology for the scenario under investigation.  In particular, the departure from alignment would allow for the $q \bar q \to Z^* \to ha$ process to occur at a non negligible level.

\begin{figure}[!h]
\vspace*{-4mm}
    \centering
    \subfloat{\includegraphics[width=0.5\linewidth]{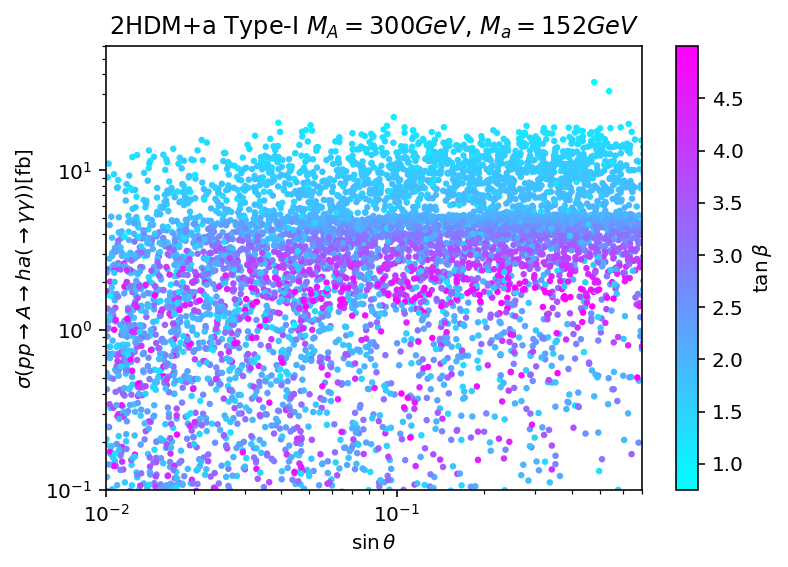}}
    \subfloat{\includegraphics[width=0.5\linewidth]{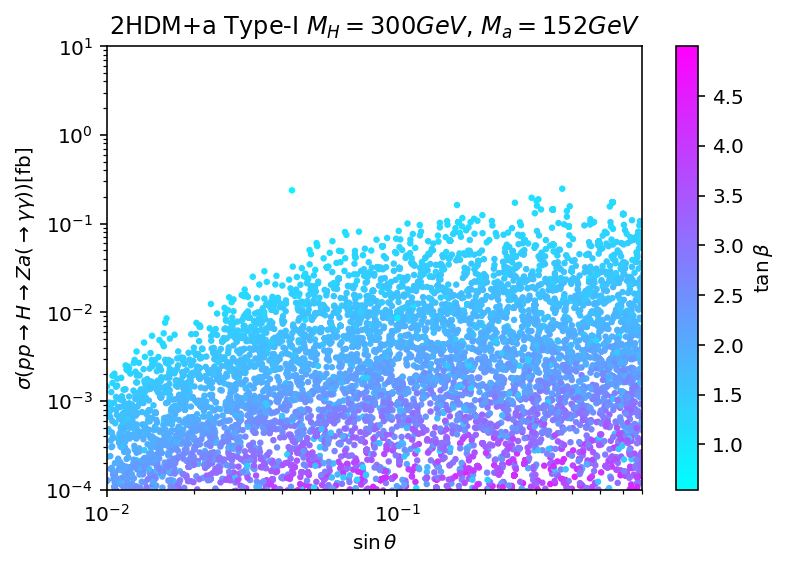}}
\vspace*{-3mm}
    \caption{\small  Values of the $pp\to A \to ah \to \bar b b \gamma \gamma$ (left)   and $pp\to H \to Za \to \bar b b \gamma \gamma$ production cross sections as a function of $\sin\theta$ for $M_a=125\,\mbox{GeV}$ in the Type-I scenario; the color code follows the value of $\tan\beta$. The two panels correspond to the fixed values $M_A=300\,\mbox{GeV}$ (left) and $M_H=300\,\mbox{GeV}$ (right).}
    \label{fig:plot_mA_300-1}
\vspace*{-3mm}
\end{figure}

We have performed two independent parameter scans of the 2HD+a model, keeping fixed alternatively $M_H$ and $M_A$ to 300 GeV and varying the other parameters in the same range as in  eq.~(\ref{eq:scan95}). The output is shown in Fig. \ref{fig:plot_mA_300-1} and it is rather similar to the one  of  Fig.~\ref{fig:95gamgam} of  the previous case. It exhibits first the cross sections times branching ratios for the processes $pp \to A\to h a \to b\bar b+\gamma\gamma$  (in the left panel)  and  $pp \to H\to  aZ \to \gamma\gamma b\bar b$ (in the right one), again as a function of $\sin\theta$ and with the corresponding value of $\tan\beta$ recorded. One can see that a cross section of about 6 fb can be obtained in the mode $pp \to A \to hZ \to \gamma\gamma b\bar b$ but barely, and only for values of $\tan\beta$ close to unity. In turn, the cross section for the companion process $pp\to H\to Za$ is way too small, at least by two order of magnitude, the reason being that the branching fraction for the $H \rightarrow Za$ is very small, as the main decay mode the $H$ boson will be into $hh$ final states. 

\begin{figure}[!h]
\vspace*{.1mm}
    \centering
\subfloat{\includegraphics[width=0.45\linewidth]{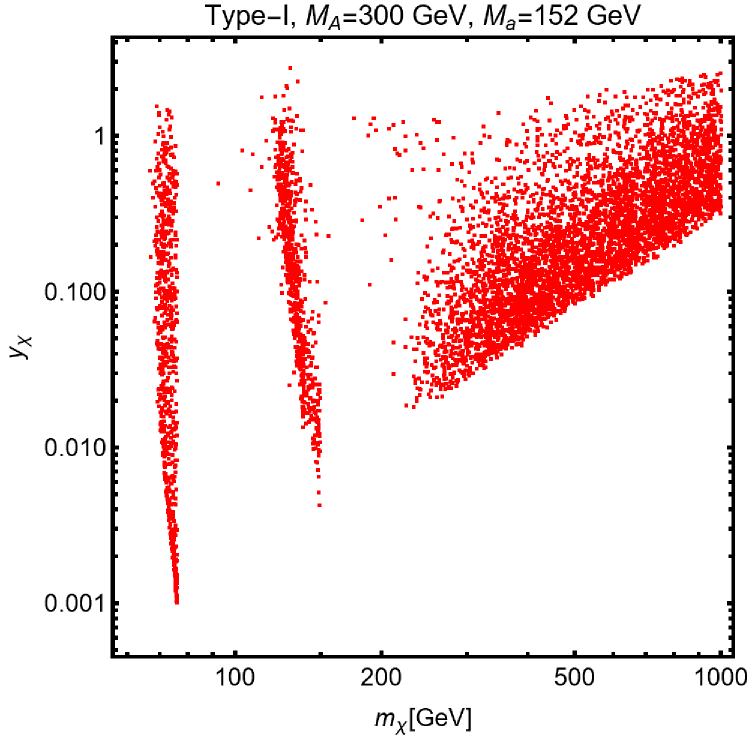}}   \subfloat{\includegraphics[width=0.45\linewidth]{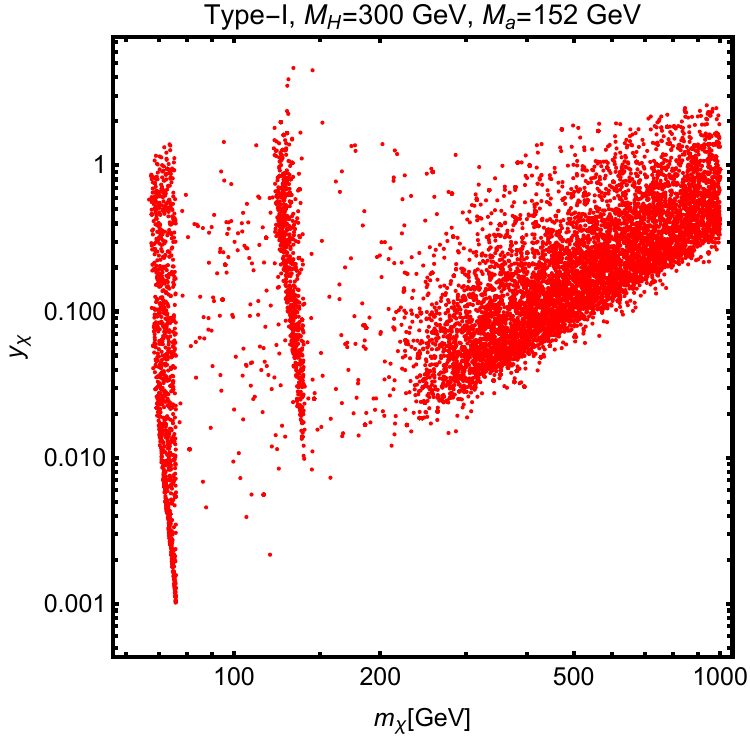}~~~}
\vspace*{-2mm}
    \caption{\small Model points complying with DM constraints in the $[m_\chi, y_\chi]$ plane in the scenario $M_a=152\,\mbox{GeV}$. The two panels refer to two distinct scans with fixed values $M_A=300\,\mbox{GeV}$ (left) and $M_H=300\,\mbox{GeV}$ (right). In both cases, the Type-I scenario has been considered.}
    \label{fig:plot_mA_300-2}
\vspace*{-5mm}
\end{figure}

Moving to the Dark Matter issue\footnote{The  DM phenomenology of the 95 GeV resonance  will be  considered at a later stage.}, we have then combined the previous scan over the parameters of the 2HD+a Higgs sector with the variation of the DM mass and couplings within the following ranges:
\begin{eqnarray}
    m_\chi \in [1,1000]\,\mbox{GeV},\,\,\,\,\,y_\chi \in [10^{-3},10] \, .
\end{eqnarray}
For each parameter assignment, we have computed the DM observables, namely relic density, direct detection cross section and present time annihilation cross section. We show in Fig. \ref{fig:plot_mA_300-2}  the fraction of the model points displayed in the previous figure, but in the $[m_\chi,y_\chi]$ bi-dimensional plane,  and complying with the limits from DM direct and indirect detection.  The two panels are associated to the two distinct scans characterized by fixed $M_A$ (left-hand panel) and $M_H$ (the right-hand one) values.

The distribution of the model points in the figure can be interpreted as follows. In the scenario $M_A=300 \,\mbox{GeV}$, for $m_\chi \leq 200 \, \mbox{GeV}$, the relic density is determined mostly by DM annihilation into SM fermions via $s$-channel exchange of the two pseudoscalar bosons. Given the $s$-wave nature of the corresponding cross section, the majority of the parameter values are ruled out by indirect detection constraints, as well as by direct detection, but to a lesser extent. The only exceptions are represented by the poles $m_\chi=\frac12 M_{a,A}$. At higher DM masses, indirect detection is not capable any longer to constrain the thermally favored value of the DM annihilation cross section and, at the same time, the very efficient annihilation process into $ha$ final states becomes kinematically accessible. 
 
 The case $M_H=300\,\mbox{GeV}$ shows a very similar pattern, the only difference being that one sees more viable parameter assignments for $\frac 12 M_a \leq m_\chi \leq \frac12 (M_h+M_a)$ and $m_\chi \neq 150\,\mbox{GeV}$. This is simply due to the fact that the value of $M_A$ can vary, moving consequently the position of the pole $m_\chi \simeq \frac12 M_A$. The larger density of points at $m_\chi \simeq 150\,\mbox{GeV}$ is due to the fact that theoretical and high-precision constraints favor the mass degeneracy $M_H=M_A$.

We should finally note, that this excess at a mass of 152 GeV, in particular in all the additional and rather involved topologies considered in Refs.~\cite{Crivellin:2021ubm,Bhattacharya:2025rfr}, can be generated by cascades of a 650 GeV Higgs as will be discussed  in subsection 3.4.\vspace*{-3mm}

\subsubsection*{3.3 A 365 GeV $\mathbf{\bar t t}$ resonance?}

Rather recently the CMS collaboration~\cite{CMS:2025kzt,CMS:2025dzq} has observed a significant excess of events,  more than  7$\sigma$, close to the $t \bar t$ threshold; an excess which has been confirmed by ATLAS shortly after \cite{ATLAS:2025kvb}. It is clear that the signal is most probably due to the widely expected production of the toponium, an $^1S_0$ quasi-bound $t \bar t$ state with a mass $m_{\eta_t} \simeq 2m_t= 345$ GeV. However, the cross section, which cannot be fully calculated using perturbative QCD  \cite{Toponium}, could well be on the high-side. Some studies, including the one performed by the CMS experiment \cite{CMS:2025kzt}, speculated about the presence of a pseudoscalar state with a mass of about 365 GeV\footnote{The experimental resolution on the $t\bar t$ invariant mass spectrum is rather large, $\Delta m_{tt} \approx 20\%~(\approx  70$ GeV in this particular case) \cite{CMS:2018cyk} allowing for this mass gap. Note that for such a mass difference, all non-perturbative and mixing effects between the toponium and the pseudoscalar boson can be ignored.} originating from a model with an extended Higgs sector (see for instance the recent Ref.~\cite{Flacke:2025dwk}), the most obvious option being the $A$ boson of a 2HDM. 

However, it has been shown (see for  instance Ref.~\cite{Lu:2024twj}) that in such a case,  one is severely constrained by both theoretical requirements and experimental data and cannot afford a large mass splitting between the $A$ and the $H,H^\pm$ states to comply with observation. In particular, the constraint $M_{H^\pm}  \gsim 800$ GeV from $b \to s\gamma$ transitions would result into a scalar potential that triggers  unitarity violation and non-perturbative Higgs self-couplings. In the rare cases in which one could have a relatively light $H$ state not degenerate in mass with $H^\pm$, constraints from LHC searches of  $t\bar t  Z$ final states, which lead to a quite copious rate in the process $gg\to H \to AZ \to t\bar t  Z$, exclude the 2HDM option \cite{Lu:2024twj}.  This is one of the reasons why a generic $A$ resonance was considered in Ref.~\cite{Djouadi:2024lyv}. 
 
In this section, we will show that as far as  the theoretical constraints are concerned, the situation is much more favorable in the 2HD+a scenario since one can have a pseudoscalar $a$ with a mass of about 350 GeV, while pushing the common mass of the 2HDM states close to the TeV scale and hence avoiding the $b \to s\gamma$ constraint. Nevertheless, in including the $A$ contribution, one should take into account the destructive interference effects between the $gg \to A \to t\bar t$ pure signal process and the QCD background $gg \to t\bar t$ which were shown to be extremely important \cite{tt-interference,Djouadi:2024lyv}. In the most recent of theses papers, it was found that for the mass and couplings to top quarks (which is mainly driven by the value of $\tan\beta$) that reproduce the early CMS signal, the large total width of a generic pseudoscalar state renders  makes it huge. For instance, if one integrates over the entire $t\bar t$ invariant mass spectrum,  one needs an initial  pure signal cross section that is about four times larger than the signal+background rate measured by CMS, namely $\sigma (tt) \approx 7$ pb for $M_A=365$ GeV. This aspect, which has not been discussed in the 2HDM \cite{Lu:2024twj} will render  the situation rather difficult even in the 2HD+a case. 

However, as alluded to earlier, there is no need to explain the entire excess observed by CMS since most of it should be attributed to the toponium resonance which is undoubtedly present.  In the following, we will again assume that the 2HDM $H,A,H^\pm$ states are too heavy, with a mass above the 800 GeV limit dictated by $b \to s\gamma$, and do not contribute and consider only the $gg \to a \to t\bar t$ process in the four Yukawa type configurations, whose rates are governed only by the parameters $\tan\beta$ and $\sin\theta$, once $M_a=365$ GeV is fixed. As our goal is simply  to estimate the possible contribution of $a$ to the signal and hence, the possible order of magnitude of the cross section, we will ignore the interference effects\footnote{As  noted,  the interference is rather large only if $a$ has enough strong couplings to reproduce the entire signal. If only a small fraction of it is targeted, say 10\% to 20\%, then the $a$ couplings and hence its total width are much smaller and would make the impact of the interference less significant impact;  for a 10\% contribution to the signal,  it might even be negligible.}.  

We have then performed a scan of the parameters of the Higgs sector over the following ranges
\begin{equation}
    M_{H,A,H^{\pm}} \in [365\,\mbox{GeV},1.5\,\mbox{TeV}],\,\,\,\, \sin\theta \in [{0.1,0.7}],\,\,\,\, \tan\beta \in [0.3,5]
\end{equation}
retaining only the parameter values complying with the constraints on the scalar potential of the model. We have computed, for the latter model points, the production cross section of the $gg \rightarrow a \rightarrow \bar t t$ process by adapting to the 2HD+a the computation for the production of the pseudoscalar A in the 2HDM provided by the SusHi package \cite{SUSHI}. Note that for the values of $\tb$ that we are considering, the decay branching fraction BR$(A \to t\bar t)$ is always close to unity once the mode is kinematically accessible. 

\begin{figure}[!h] 
\vspace*{-3mm}
    \centering
\subfloat{\includegraphics[width=0.5\linewidth]{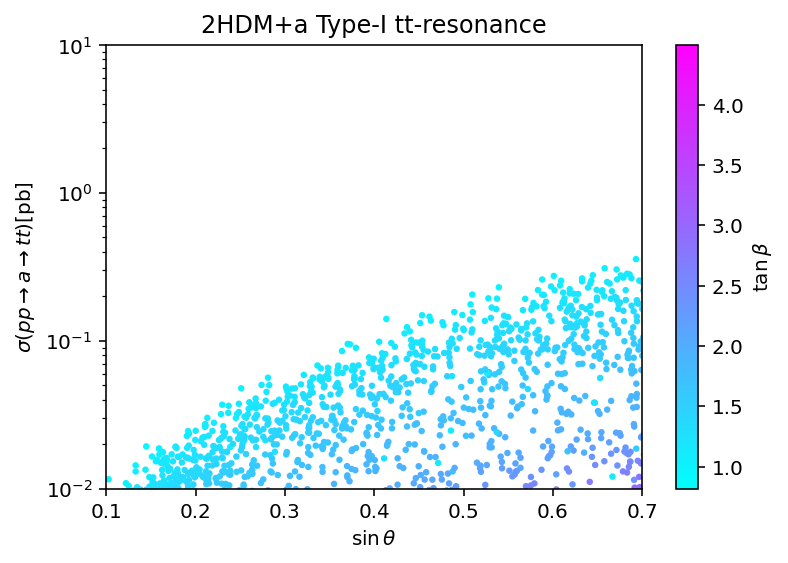}}
    \subfloat{\includegraphics[width=0.5\linewidth]{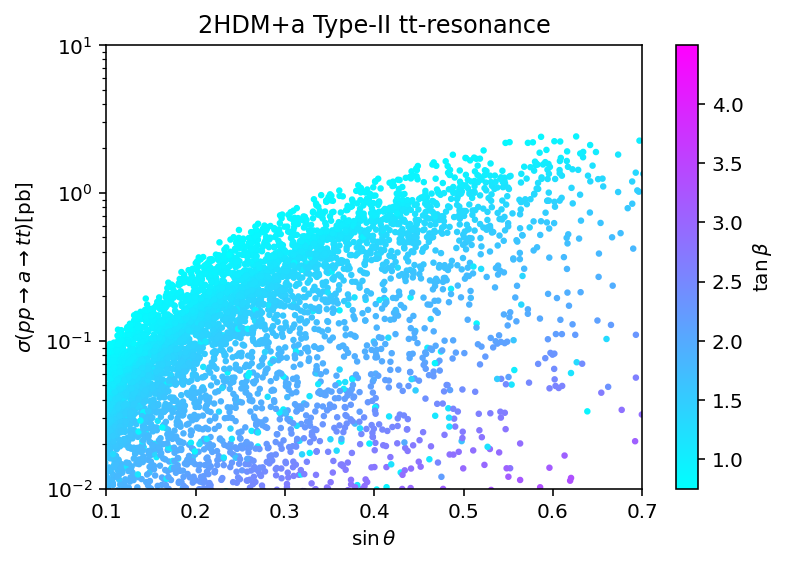}}\\[-3mm]
     \subfloat{\includegraphics[width=0.5\linewidth]{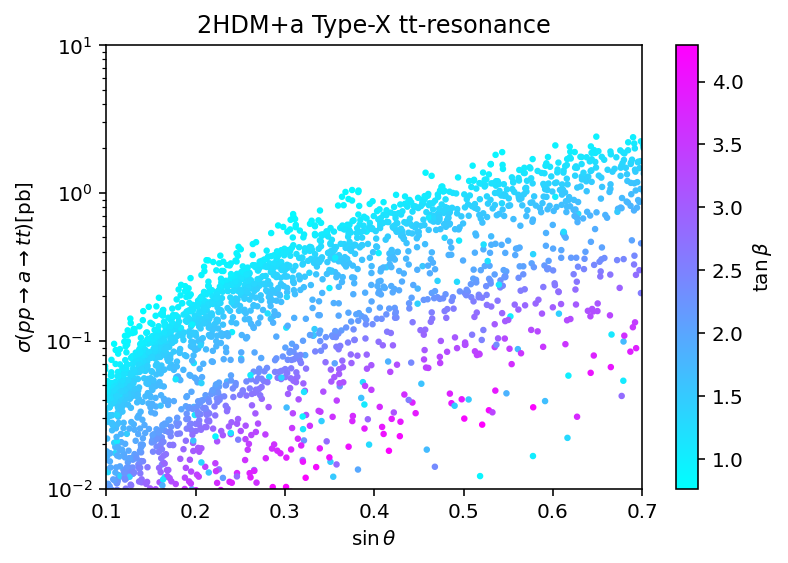}}
    \subfloat{\includegraphics[width=0.5\linewidth]{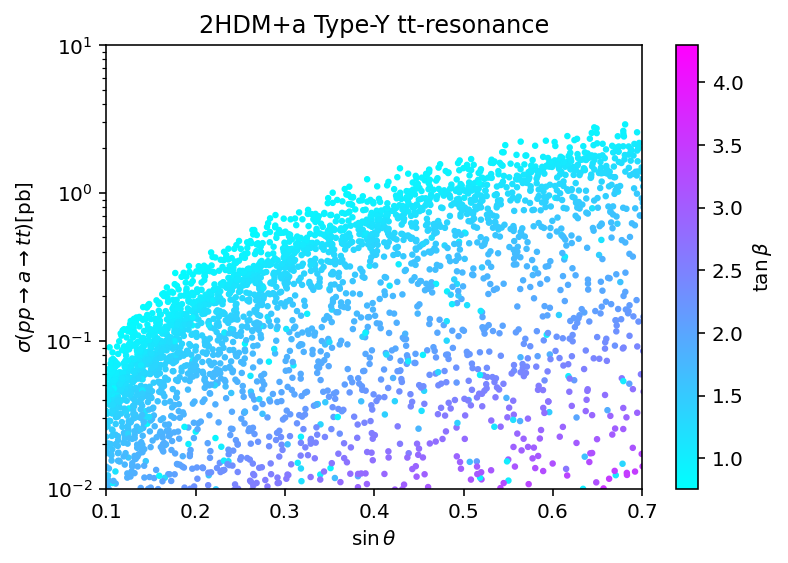}}
\vspace*{-3mm}
    \caption{\small Production cross section of a 365 GeV resonance, interpreted as the light pseudoscalar $a$ of the 2HD+a model, decaying into top pairs, as function of $\sin\theta$. The points of the scatter plot corresponds to parameter assignments of the model complying with theoretical and experimental constraints; the color pattern follows the value of $\tan\beta$. The different panels correspond the four types of Yukawa configurations reported on top of each plot.}
    \label{fig:scantt}
\vspace*{-3mm} 
\end{figure}

As we have done earlier, we show in the four panels of Fig.~\ref{fig:scantt},  corresponding to the four Yukawa configurations, namely Type-I, Type-II, Type-X and Type-Y, the value of the $gg \to a \to \bar t t$ production cross section, as function of $\sin\theta$. The model points passing the constraints are displayed according to a color code that follows the value of $\tan\beta$. As can be seen, it is very difficult to achieve a cross section times branching fractions of about 7 pb, which is the amount of signal observed by CMS; only maximal values of the mixing angle in Type-II and Type-Y (which approximately give the same results) allow for it. This result can be explained as follows. In the conventional 2HDM setup, the needed value of the $\bar t t$ production cross section is achieved for $\tan\beta=1.28$~\cite{Lu:2024twj}. Even if interpreting the resonance with the light pseudoscalar state $a$ allows to relax many of the constraints disfavoring the interpretation in the conventional 2HDM, the $\sin\theta$ suppression of the couplings of $a$ requires values of $\tan\beta<1$ to fit the excess, putting the 2HD+a interpretation in tension with the limits from $B_s \rightarrow \mu^+ \mu^-$ as discussed previously. However, as discussed earlier, since the existence of the  toponium  $\eta_t$ resonance is firmly established, only a fraction of this rate could eventually be attributed to the $a$ resonance. As can be seen, an approximate 20\% contribution is achievable in  a large portion of the parameter space.

Again, we have  confronted the results of the collider study with the solution of the DM puzzle using the same procedure illustrated in the previous subsection.  As discussed in detail elsewhere, see for example Ref.~\cite{2HDa-us}, the DM phenomenology of the Type-I (Type-II) and Type-X (Type-Y) are very similar and, in fact, if one ignores all Yukawa Higgs couplings except for the ones involving the top quarks,  it is true for the four configurations and almost all aspects of the model. Consequently, we will simply show explicitly, the numerical results relative to the Type-I and Type-II scenarios which are given in Fig. \ref{fig:scantt}. Displayed  in the $[m_\chi,y_\chi]$ bi-dimensional plane, the model points that account for the correct DM relic density and, at the same time, are compatible with present bounds from direct and indirect detection.

\begin{figure}[!h]
\vspace*{-3mm}
    \centering
    \includegraphics[width=0.45\linewidth]{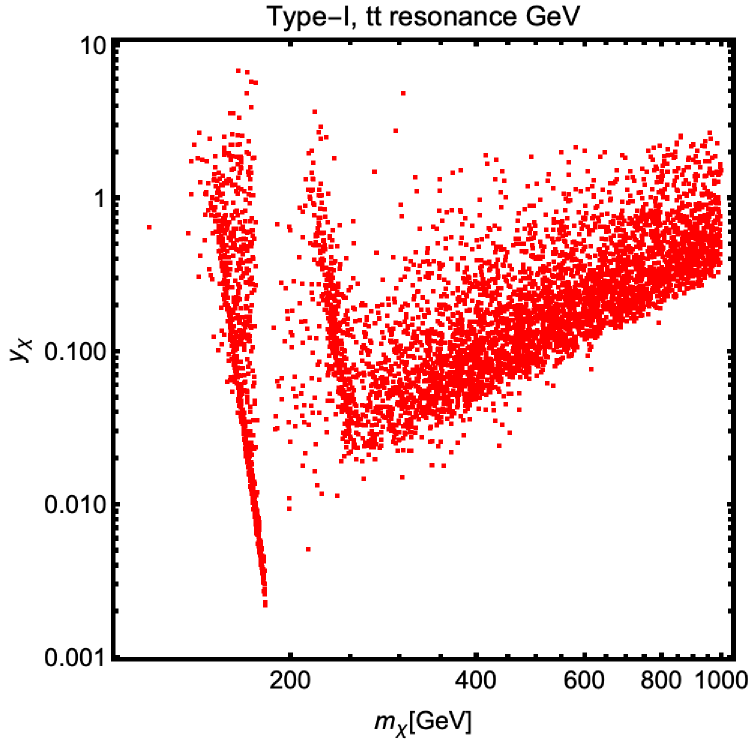}
    \includegraphics[width=0.45\linewidth]{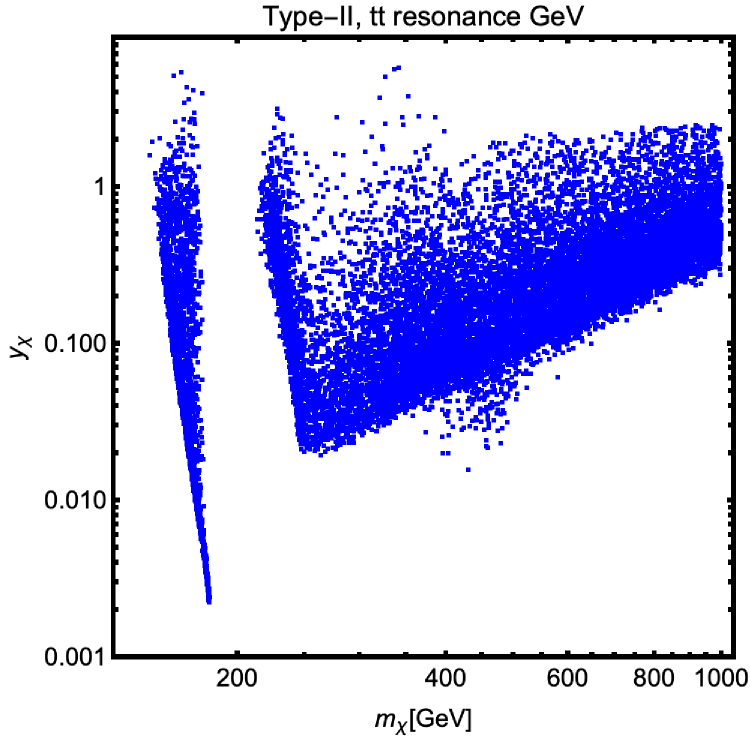}
\vspace*{-3mm}
    \caption{\small 2HD+a configurations with $M_a=365\,\mbox{GeV}$ complying with DM constraints
    in the $[m_\chi,y_\chi]$ plane. The constraints are shown for the Type I (left) and Type II (right).}
    \label{fig:plotttDM}
    \vspace*{-3mm}
\end{figure}

The distribution of model points bears many similarities with what was already shown in Fig. \ref{fig:plot_mA_300-2}. Again, as long as the $ha$ final state is not kinematically accessible, the experimental constraints are mostly fulfilled for $m_\chi \simeq \frac12 M_a$. On the contrary, vast regions of the parameter space become viable for the DM mass range $m_\chi > \frac 12 (M_h+M_a) \simeq 245\,\mbox{GeV}$.\vspace*{-3mm}

\subsubsection*{3.4 A 650 GeV resonance?}

A detailed discussion of the various excesses which can be attributed to a putative resonance of 650 GeV can be found, for instance, in Ref.~\cite{LeYaouanc:2025mpk} together with some interpretation in various new physics models (like the 2HDM, a triplet Higgs and even the Georgi Machacek or extra-dimensional models) and in terms of various cascade decays involving some of the resonance signals that we have already discussed\footnote{In addition to the resonant 95, 152 and 350 GeV final states that we have just studied, Ref.~\cite{LeYaouanc:2025mpk} invokes also a possible signal for a pseudoscalar Higgs boson with a mass of 470 GeV observed by the ATLAS collaboration \cite{ATLAS:2022fpx}, in particular in the $hhZ \to bbbb\ell\ell$ final state. }. In this work, we will discuss an interpretation, within the 2HD+a model, of the signature studied in Ref.~\cite{CMS:2023boe}, namely of the type $pp \rightarrow X \rightarrow Y(\rightarrow \bar b b) h (\rightarrow \gamma \gamma)$. As pointed out in the relevant reference, the largest deviation from the SM background (yet not conclusive) is achieved for invariant masses $M_X=650\,\mbox{GeV}$ and $M_Y=90\,\mbox{GeV}$. It is interesting to investigate the potential correlation between this $\bar b b \gamma \gamma$ signal and the diphoton excess discussed at the beginning of the section. 

Within our 2HD+a setup, the $X$ and $Y$ states are interpreted as being, respectively, the $A$ and $a$ pseudoscalar states. The $Z$ boson could be, alternatively, identified with the $Y$ state. However, the $pp \rightarrow A \rightarrow Zh$ mode has a more suppressed rate than $pp \rightarrow A \rightarrow ah$, since the $AZh$ coupling is proportional to the deviation from the alignment limit. As customary, we have performed in the four Yukawa coupling configurations, a parameter scan within the ranges reported in  eq.~(\ref{eq:scan95}) but fixing $M_A=650\,\mbox{GeV}$.

\begin{figure}[!h] 
\vspace*{-.1mm}
    \centering
    \includegraphics[width=0.49\linewidth]{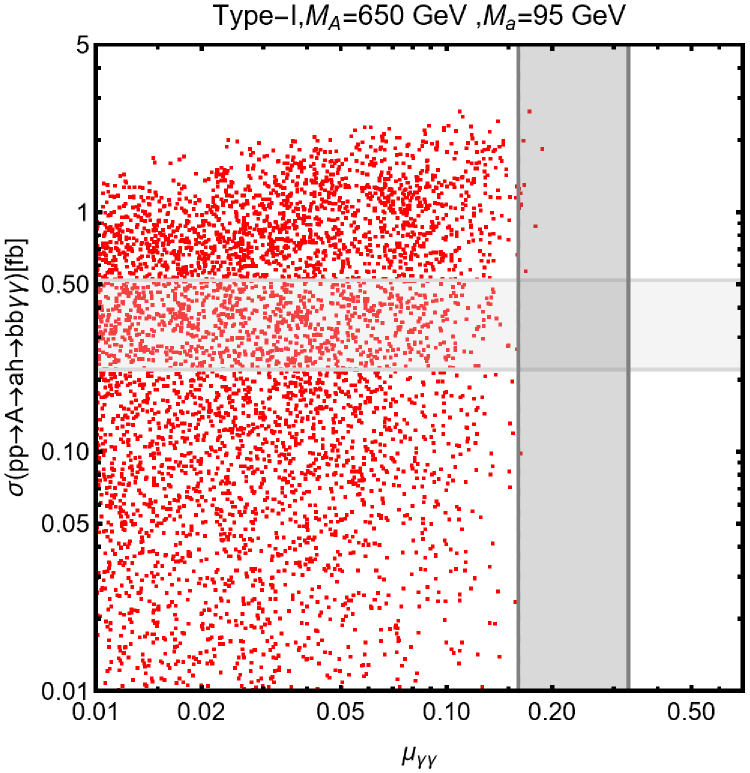}
    \includegraphics[width=0.49\linewidth]{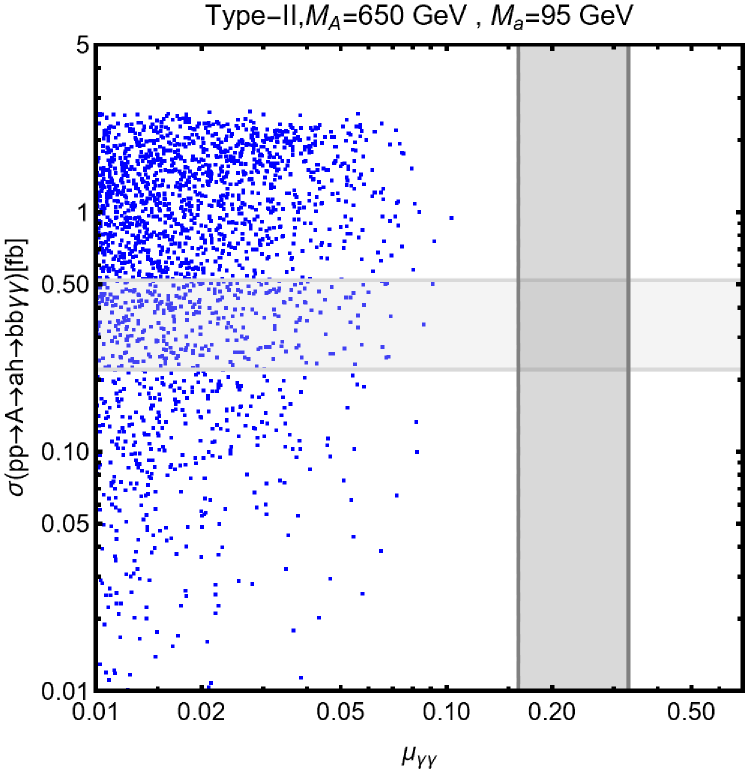}\\[1mm]
    \includegraphics[width=0.49\linewidth]{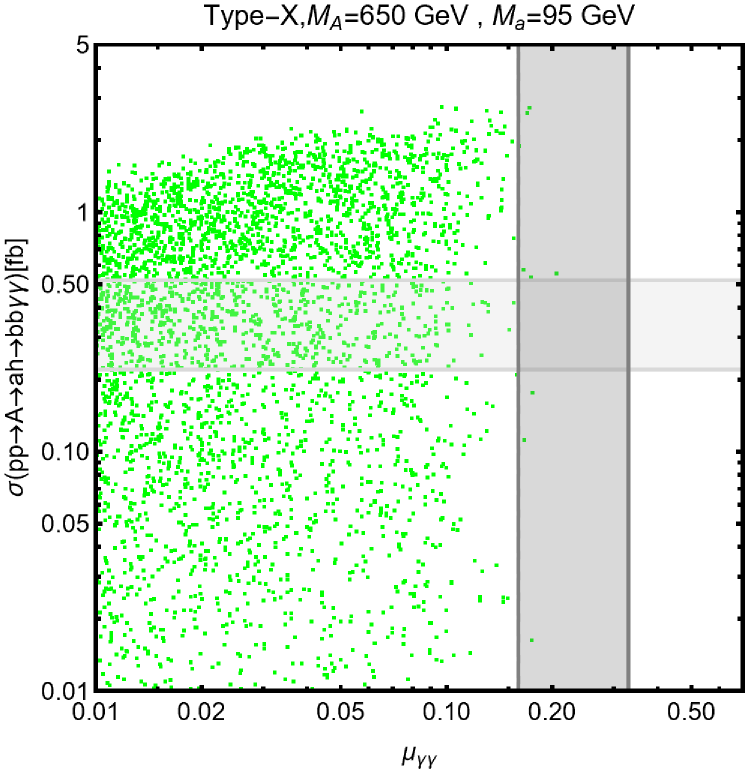}
    \includegraphics[width=0.49\linewidth]{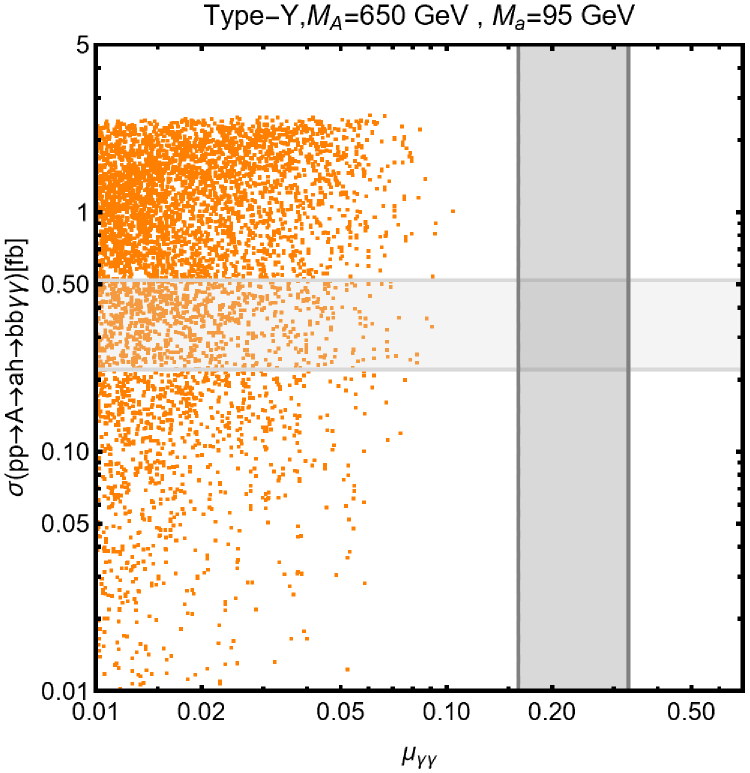}
    \caption{\small Outcome of a parameter scan of the 2HD+a model assuming that the heavy and light pseudoscalars have masses of respectively of 650 GeV and 95 GeV in the four Yukawa configurations. In each panel, the $x$-axis is for the signal strength of $a\to \gamma\gamma$ with $M_a=95$ GeV, while the $y$-axis represents the cross section for $pp\rightarrow A\rightarrow a(\rightarrow \bar b b) h (\rightarrow \gamma \gamma)$. The gray bands represent the values of the observables corresponding to the experimental excesses within  $2\sigma$.}
    \label{fig:p650fit}
\vspace*{3mm}
\end{figure}

The outcome is presented in the four panels of Fig. \ref{fig:p650fit}, representing the four Yukawa coupling configurations. The $x$-axis is for the signal strength $\mu_{\gamma \gamma}$ of the 95 GeV resonance previously defined in eq. (\ref{eq:mugamgam}),  while the $y$-axis is for the cross section of the process $pp \rightarrow A \rightarrow ah$. We have verified that in the current setup, this process accounts for most of the $\bar b b \gamma \gamma$ signal as the $A \rightarrow Zh$ decay would require a large deviation from the alignment limit to have a significant branching fraction. The horizontal and vertical gray bands represent, respectively, the value $\sigma (pp \rightarrow \bar b b \gamma \gamma)=0.35^{+0.17}_{-0.13}\,\mbox{fb}$ corresponding to the excess reported by CMS \cite{Benbrik:2025hol} and the value $\mu_{\gamma \gamma}$ compatible, within $2 \sigma$, with the 95 GeV excess. 

As one can imagine, a combined fit of the two  signals, corresponding to the case in which points of the scatter plot would fall at the intersection of the two bands, is  complicated in general. In particular, the Type-II and Type-Y setups have strong difficulties to comply with the 95 GeV excess, in particular because the required value $M_A=650 \,\mbox{GeV}$ will conflict with the constraint  $M_{H^{\pm}} \geq 800 \,\mbox{GeV}$  from $b \rightarrow s \gamma$ transitions. In the case of the Type-I and Type-X, even if there are no scan-points that lie at the intersection of the bands, a marginal possibility for a combined fit of the signals associated with the 650 GeV and 95 GeV resonances might emerge from either,  a more sophisticated numerical analysis or, by extending to 3$\sigma$ the fit intervals for the two observables displayed in the figure.

\begin{figure}[!h] 
\vspace*{1mm}
    \centering
    \includegraphics[width=0.45\linewidth]{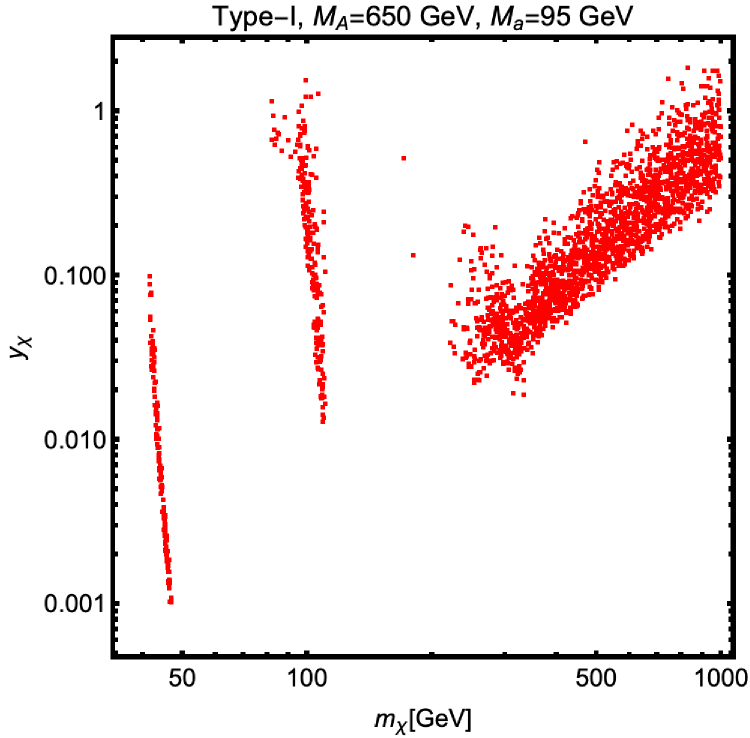}
    \includegraphics[width=0.45\linewidth]{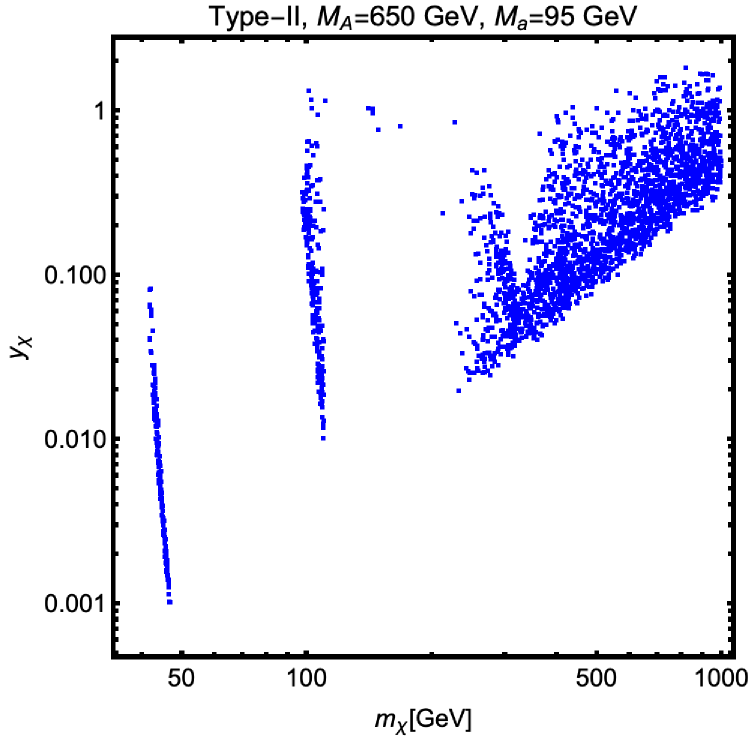}\\
\vspace*{-.2mm}
    \caption{\small Model points of the 2HD+a  $[m_\chi,y_\chi]$ bidimensional plane in the case $M_a=95 \,\mbox{GeV}$ and $M_A=650\,\mbox{GeV}$ and complying with DM constraints. We have restricted the Type-I (left) and Type-II (right) configurations.}
    \label{fig:p650DM}
\vspace*{-1mm}    
\end{figure}

As already done before, we have complemented our phenomenological analysis with the aspects related to DM. We have again considered the coupling of the Higgs sector with a fermionic singlet DM and computed relic density, direct and indirect detection  cross sections by freely varying the DM mass and couplings. Fig.~\ref{fig:p650DM} shows, in the conventional $[m_\chi,y_\chi]$ bidimensional plane the points complying with all these constraints, considering only the Type-I and Type-II Yukawa configurations. For $m_\chi \leq \frac12 (M_h+M_a)$, the viable model points lie in the vicinity  of the $m_\chi \simeq \frac 12 M_a$ "pole". In the mass range $m_\chi \simeq O(100\!-\!200)\,\mbox{GeV}$,  most of the model points are excluded by indirect detection, with the exception of a tiny strip in proximity of the kinematical threshold of the annihilation channel into $ha$. Hence, heavy DM candidates are favored, with mass above 200 GeV.

\begin{figure}[!h] 
\vspace*{-3mm}
    \centering
    \includegraphics[width=0.49\linewidth]{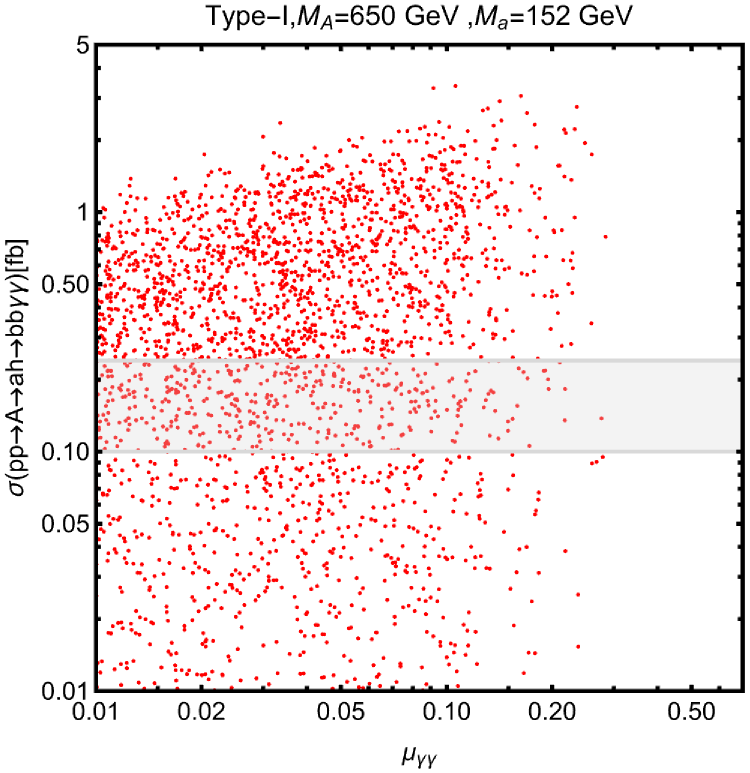}
    \includegraphics[width=0.49\linewidth]{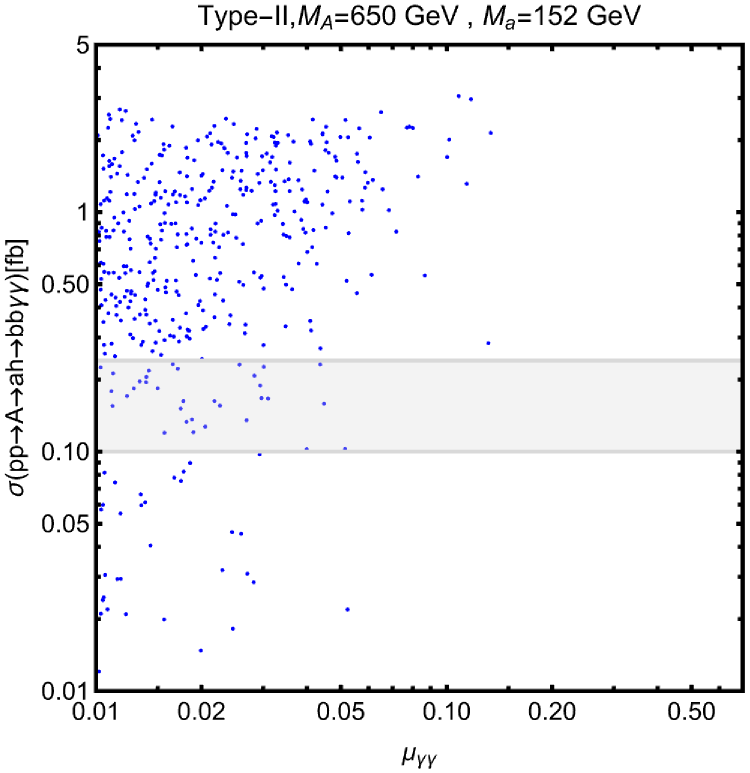}\\[1mm]
    \includegraphics[width=0.49\linewidth]{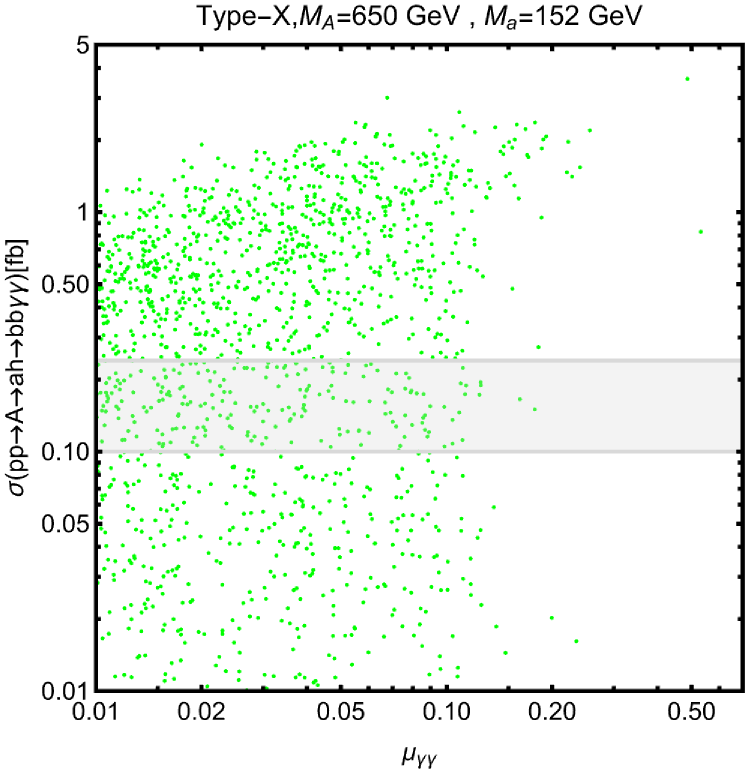}
    \includegraphics[width=0.49\linewidth]{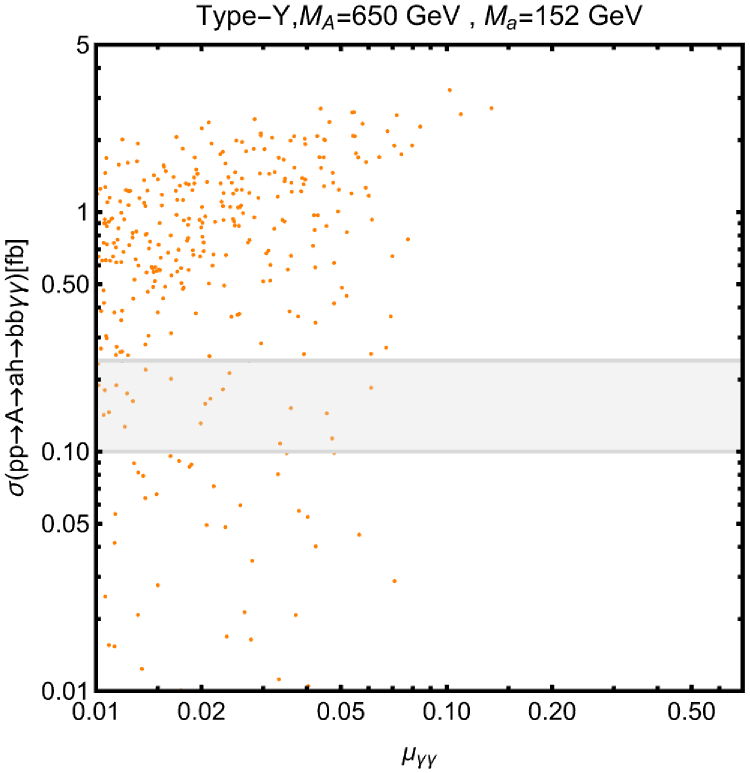}
    \caption{\small Analogous plot as Fig. \ref{fig:p650fit} but considering $M_A=650\,\mbox{GeV}$ and $M_a=152\,\mbox{GeV}$. The gray band represents again, within $2\sigma$, the signal region as given in \cite{CMS:2023boe}.}
    \label{fig:p650fit152}
\vspace*{-3mm}
\end{figure}

Finally, a last scenario that we will briefly discuss is the case in which a 650 GeV heavy resonance decays into a 152 state and the SM Higgs boson leading, again, to the $\bar b b \gamma \gamma$ topology. For this purpose, we have repeated the previous parameter scan in  the four Yukawa configurations, fixing this time $M_a=152\,\mbox{GeV}$ while we still keep $M_A=650$ GeV. We  display in Fig. \ref{fig:p650fit152}  the results in the $[\mu_{\gamma \gamma}, \sigma (pp\rightarrow A\rightarrow ah \rightarrow \bar b b \gamma \gamma)]$ plane. In each panel, the horizontal gray band corresponds to the value $\sigma (pp \rightarrow \bar b b \gamma \gamma)=0.15^{+0.09}_{-0.05}\,\mbox{fb}$ \cite{CMS:2023boe} and one can see that many points fall in the requested interval. We have not explicitly reported, for this scenario, results related to DM phenomenology as the outcome would be totally analogous to the previous case, apart the different position of the $\frac12 M_a$ "pole".  

\subsection*{4. Conclusions} 

In this paper, we have endeavored  to interpret, in the context of a model with two-Higgs doublet plus a light singlet pseudoscalar Higgs state or 2HD+a model, the various excesses that seem to  have appeared in LHC data in the search for additional Higgs bosons performed by the ATLAS and CMS collaborations, beyond the one which was discovered a decade ago. Besides the observation of a toponium resonance  in $t\bar t$ events near the 350 GeV threshold, which was largely expected in QCD but which could be also accompanied by a new Higgs particle with a comparable mass and decaying into $t\bar t$ pairs, we have  considered various modest excesses  involving a new resonance decaying into  two photons accompanied  eventually with some other particles like $b\bar b$ pairs or missing energy. These occurred at masses of 95 GeV, 152 GeV as well as at 650 GeV, values which were taken to be the  mass of the lightest $a$ boson in the first two cases and, in the latter case, that of the  heavier 2HDM neutral  state  decaying through cascades into the $a$ boson. These excesses are rather difficult to interpret in the most favored new physics extensions like the 2HDM or MSSM, and more involved scenarios need to be considered. 

After performing detailed scans of the relatively large parameter space of our 2HD+a model, and taking into account all possible constraints, the theoretical ones from unitarity and perturbativity as well as the experimental ones, in particular those set by flavor physics and other LHC searches and measurements, we have shown that there are still values of the Yukawa coupling of the $a$ state to fermions and of its mixing with the heavy 2HDM state $A$, for which these excesses at the different invariant mass values could be indeed explained. The specific range of the latter  parameters still allows to achieve the correct cosmological relic density for the additional dark matter particle of the model, and to be within the limits set by direct and indirect searches for it in astro-particle physics experiments. 

We did not attempt to perform a detailed simulation of all the signal events, taking into account all topologies and the experimental conditions. As we noted before, most of thee excesses are not firm and their specific features not well established. Our main goal was therefore simply to show that there are still possibilities to interpret them in our 2HD+a scenario without being in conflict with the relevant theoretical and experimental constraints. In the fortunate case where one of these signals becomes firm and turns to a discovery, a more detailed analysis would have then to be performed.\bigskip

\noindent {\bf Acknowledgements}:\\  
The work of A.D. was supported by the  Spanish grant PID2024-161668NB-I00.


\begin{thebibliography}{999}
 
\bibitem{Hdiscovery} 
ATLAS Collaboration, Phys. Lett. B716 (2012) 1–29;  
CMS Collaboration, Phys. Lett. B716 (2012) 30–61. 

\bibitem{DM-reviews} For recent reviews on Dark Matter, see e.g. M.~Cirelli, A.~Strumia and J.~Zupan, arXiv:2406.01705; 
G.~Arcadi et al. Eur. Phys. J. C78 (2018) no.3, 203;   Eur. Phys. J. C85 (2025) 2, 152. 

\bibitem{H-portal} G.~Arcadi, A.~Djouadi and M.~Raidal, Phys. Rept. 842 (2020) 1.

\bibitem{HHunting} ATLAS and CMS Collaboration we sites,\\
{\tt https://twiki.cern.ch/twiki/bin/view/AtlasPublic/HDBSPublicResults};\\
{\tt https://cms-results.web.cern.ch/cms-results/public-results/ \\ publications/HIG/SUS.html}. 
For summaries, see; Kerstin Tackmann, Experimental Summary talk at Higgs Hunting, September 2024; Maria Cepeda, idem, July 2025. 

\bibitem{ParticleDataGroup:2024cfk} S. Navas et al., Particle Data Group, Phys. Rev. D110 (2024) 3, 030001. 

\bibitem{Muong-2:2025xyk} D.P. Aguillard, Muon g-2 collaboration, Phys. Rev. Lett. 135, 10 (2025) 101802. 

\bibitem{Flavor-review}  Y.S. Amhis et al., Heavy Flavor Averaging Group, Phys. Rev. D 107, 052008 (2023).

\bibitem{LZ:2024zvo} J. Aalbers, LZ collaboration, Phys. Rev. Lett. 135, 1 (2025) 011802. 

\bibitem{McDaniel:2023bju} A.~McDaniel et al.,  Phys. Rev. D109 (2024) no.6, 063024.

\bibitem{2HDM} For a review on 2HDMs, see G. Branco et al.,  Phys. Rept. 516 (2012) 1. 

\bibitem{MSSM} For a review of the MSSM Higgs sector, see A.~Djouadi,  Phys. Rept. 459 (2008) 1. 

\bibitem{2HDa-all} 
S. Ipek, D. McKeen and A.E. Nelson, Phys. Rev. D90, 5 (2014) 055021;
M. Bauer, U. Haisch and F. Kahlhoefer, JHEP 05 (2017) 138; 
P. Tunney, J.M.No and M. Fairbairn, Phys. Rev. D96, 9 (2017) 095020, 2017;
T. Robens. The THDMa Revisited. Symmetry, 13(12) (2021) 2341;
G.~Arcadi and A.~Djouadi, Phys. Rev. D106 (2022) no.9, 095008; 
S. Argyropoulos and U. Haisch,     SciPost Phys. 13 (2022) 1, 007;
G.~Arcadi, A.~Djouadi and F.~d.~Queiroz, Phys. Lett. B834 (2022) 137436.

\bibitem{2HDa-pot}  T. Abe et al., LHC Dark Matter Working Group, Phys. Dark Univ. 27 (2020) 100351.

\bibitem{Goncalves:2016iyg}  D.~Goncalves, P.A.N.~Machado and J.M.~No, Phys. Rev. D95 (2017) no.5, 055027. 
 
\bibitem{2HDa-us} G.~Arcadi, N.~Benincasa, A.~Djouadi and K.~Kannike, Phys. Rev. D108 (2023) no.5, 055010.

\bibitem{CMS:2018cyk} A.~M.~Sirunyan et al., CMS collaboration, Phys. Lett. B 793 (2019), 320. 

\bibitem{CMS:2024yhz} 
A.~M.~Sirunyan et al. CMS collaboration, Phys. Lett. B \textbf{860} (2025), 139067. 

\bibitem{ATLAS:2024bjr}
 G.~Aad, ATLAS collaboration, 
JHEP \textbf{01} (2025), 053.

\bibitem{Biekotter:2023oen} T. Biek{\"o}tter, S. Heinemeyer and G. Weiglein, Phys. Rev. D109, 3 (2024) 035005. 

\bibitem{CMS:2023boe} A.~Tumasyan et al., CMS collaboration, JHEP 05 (2024), 316.

\bibitem{Crivellin:2021ubm} A. Crivellin et al.,  Phys. Rev. D108, 11 (2023) 115031. 

\bibitem{Arcadi:2023smv} G.~Arcadi, G.~Busoni, D.~Cabo-Almeida and N.~Krishnan, Phys. Rev. D110 (2024) no.11, 11. 

\bibitem{Chen:2025vtg} T.K.~Chen, C.W.~Chiang, S.~Heinemeyer and G.~Weiglein, arXiv:2511.04796 [hep-ph].

\bibitem{Bhattacharya:2025rfr} S.~Bhattacharya et al.,  arXiv:2503.16245 [hep-ph];
S. Ashanujjaman et al.,  Phys. Lett. B862  (2025) 139298; JHEP 04 (2025) 003; S. Bhattacharya et al., arXiv:2306.17209; 
S. Banik et al.,  arXiv:2412.00523;
A.~Kundu et al.,  arXiv:2211.11723.

\bibitem{LeYaouanc:2025mpk} A.~Le Yaouanc and F.~Richard, New resonances at LHC, arXiv:2506.09490 [hep-ph].

\bibitem{Excesses-NMSSM}
U.~Ellwanger and C.~Hugonie, Eur. Phys. J. C \textbf{83} (2023) no.12, 1138;
U.~Ellwanger, C.~Hugonie, S.~F.~King and S.~Moretti, Eur. Phys. J. C \textbf{84} (2024) no.8, 788;
J. Cao et al., Phys. Rev. D101 (2020) 055008;  
W. Li, H. Qiao and J. Zhu, Chin. Phys. C 47 (2023) 123102; 
J. Cao et al.,  Phys. Rev. D 109 (2024) 075001;
J. Lian,  Phys. Rev. D 110 (2024) 115018;
 J. Cao et al.,  Phys. Rev. D 110 (2024) 115039; 
J. Lian and Y.-B. Liu, arXiv:2511.04968. 

\bibitem{All-papers} 
S. Ashanujjaman et al.,  Phys. Rev. D 108 (2023) L091704; 
J.A. Aguilar-Saavedra and F.R. Joaquim, Eur. Phys. J. C 80 (2020) 403;
A. Kundu, S. Maharana and P. Mondal, Nucl. Phys. B 955 (2020) 115057;
P.J. Fox and N. Weiner, JHEP 08 (2018) 025;
A. Belyaev et al.,  JHEP05 (2024) 209;
D. Azevedo, T. Biekotter and P.M. Ferreira, JHEP 11 (2023) 017;
R. Benbrik, M. Boukidi and B. Manaut, Nucl. Phys. B 1005 (2024) 116593;
R. Benbrik et al., Phys. Lett. B 832 (2022) 137245;
U. Haisch and A. Malinauskas, JHEP03 (2018) 135;
T. Biekotter, M. Chakraborti and S. Heinemeyer, Eur. Phys. J. C 80 (2020) 2; Int. J. Mod. Phys. A 36 (2021) 2142018;
T. Biek¨otter and M.O. Olea-Romacho, JHEP 10 (2021) 215;
S. Heinemeyer et al., Phys. Rev. D 106 (2022) 075003; 
T. Biekootter, S. Heinemeyer and G. Weiglein, JHEP 08 (2022) 201; Phys. Lett. B 846 (2023) 138217;  
J.A. Aguilar-Saavedra et al., arXiv:2307.03768;
S. Banik et al., Phys. Rev. D 108 (2023) 075011;
J. Dutta et al., Eur. Phys. J. C 84 (2024) 926;
D. Sachdeva and S. Sadhukhan, Phys. Rev. D 101 (2020) 055045;
R. Vega, R. Vega-Morales and K. Xie, JHEP 06 (2018) 137; 
D. Borah et al., Phys. Rev. D 109 (2024) 055021; 
A. Ahriche, Phys. Rev. D 110 (2024) 035010;
T.-K. Chen et al., Phys. Rev. D 109 (2024) 075043;
P.S.B. Dev, R.N. Mohapatra and Y. Zhang, Phys. Lett. B 849 (2024) 138481;
K. Wang and J. Zhu, Chin. Phys. C 48 (2024) 073105; 
A. Ahriche et al.,  Phys. Rev. D 110 (2024) 015025;
C.-X. Liu et al.,  Phys. Rev. D 109 (2024) 056001;
H. Xu et al.,  arXiv:2505.03592;
G. Abbas, V. Singh and N. Singh,  arXiv:2504.21593; 
Z. Li, N. Liu and B. Zhu, arXiv:2504.21273;
X. Du, H. Liu and Q. Chang, Phys. Rev. D112 (2025) 015019;
S. Gao et al., Nucl. Phys. B 1018 (2025) 117026; 
J. Gao et al., Phys. Rev. D 110 (2024) 115045;
R. Benbrik et al.,  Phys. Lett. B 868 (2025) 139688.

\bibitem{CMS:2025kzt} A.~Hayrapetyan et al., CMS  collaboration, Rept. Prog. Phys. 88 (2025) no.8, 087801.

\bibitem{CMS:2025dzq} A.~Hayrapetyan et al., CMS  collaboration, arXiv:2507.05119 [hep-ex].

\bibitem{ATLAS:2025kvb} The ATLAS collaboration, ATLAS-CONF-2025-008, July 2025. 

\bibitem{Toponium} 
B.~Fuks, K.~Hagiwara, K.~Ma and Y.~J.~Zheng, Phys. Rev. D104 (2021) no.3, 034023; Eur. Phys. J. C 85 (2025) 157; 
J.~A.~Aguilar-Saavedra, Phys. Rev. D110 (2024) no.5, 054032;
S.J.~Jiang, B.Q.~Li, G.Z.~Xu and K.Y.~Liu, arXiv:2412.18527; 
M.~V.~Garzelli et al., Phys. Lett. B866 (2025) 139532. 
 
 \bibitem{Djouadi:2024lyv} A. Djouadi,  J. Ellis and J. Quevillon, Phys. Lett. B866 (2025) 139583. 
 
\bibitem{Lu:2024twj} C.T.~Lu, K.~Cheung, D.~Kim, S.~Lee and J.~Song,  Phys. Lett. B859 (2024) 139121. 


\bibitem{Aliberti:2025beg}  R. Aliberti et al.,    Phys. Rept. 1143 (2025) 1.

\bibitem{bsgamma} M. Misiak and M. Steinhauser,  Eur. Phys. J., C77(3):201, 2017. 

\bibitem{bsmumu} 
P.~Arnan, D.~Be{\v{c}}irevi{\'c}, F.~Mescia and O.~Sumensari,
Eur. Phys. J. C{77} (2017) no.11, 796.


\bibitem{alignment} See, e.g., M. Carena et al., JHEP 1404 (2014) 015; 
J.~Bernon et al., Phys. Rev. D92 (2015) no.7, 075004;
A. Penuelas and A. Pich, JHEP 12 (2017) 084. 

\bibitem{Deltarho} J. Haller et al., Eur. Phys. J.C 78 (2018) 8, 675; J. de Blas et al.,   Phys. Rev. D106 (2022) 3, 033003.

\bibitem{Kanemura:2004mg}  S. Kanemura, Y. Okada, E. Senaha and C.P. Yuan, Phys. Rev. D70 (2004) 115002. 

\bibitem{Arcadi:2020gge} G. Arcadi, G. Busoni, T. Hugle and V. Titus,  JHEP 06 (2020) 098. 

\bibitem{h-strengths} The ATLAS experiment, Nature, 607 (7917):52–59, 2022; 
the CMS experiment,  Nature, 607(7917):60–68, 2022.


\bibitem{Enomoto:2015wbn} T.~Enomoto and R.~Watanabe, JHEP {05} (2016) 002.

\bibitem{Cheng:2015yfu} X.~D.~Cheng, Y.D.~Yang and X.B.~Yuan, Eur. Phys. J. C{76} (2016) no.3, 151.

\bibitem{HFLAV:2022esi} Y.~S.~Amhis et al. [HFLAV group], Phys. Rev. D{107} (2023) no.5, 052008.

\bibitem{Abe:2015oca} T.~Abe, R.~Sato and K.~Yagyu, JHEP 07 (2015) 064. 

\bibitem{Chun:2016hzs} E.J.~Chun and J.~Kim, JHEP 07 (2016) 110.

\bibitem{CMS:2019buh} A.M.~Sirunyan et al., CMS collaboration, Phys. Rev. Lett. 124 (2020) no.13, 131802.

 \bibitem{Flacke:2025dwk} T.~Flacke et al., arXiv:2512.03220 [hep-ph].

\bibitem{tt-interference}  K.J.F. Gaemers and F. Hoogeveen, Phys. Lett. 146B (1984) 347; D. Dicus, A. Stange and S. Willenbrock, Phys. Lett. B 333 (1994) 126; S. Moretti and D.A. Ross, Phys. Lett. B712 (2012) 245; R. Frederix and F. Maltoni, JHEP 01 (2009) 047; R. Barcelo and M. Masip, Phys. Rev. D81 (2010) 075019; A. Djouadi, R. Godbole, J. Ellis and J. Quevillon,  JHEP 03 (2016) 205; A. Djouadi, J. Ellis, and J. Quevillon, JHEP 07 (2016) 105; A. Djouadi, J. Ellis, A. Popov and  J. Quevillon, JHEP 03 (2019) 119; F.~Maltoni, C.~Severi, S.~Tentori and E.~Vryonidou, JHEP (2024) 099; H.~Bahl, R.~Kumar and G.~Weiglein, JHEP 05 (2025) 098.

\bibitem{SUSHI} R. Harlander, S. Liebler and H. Mantler, Comp. Phys. Comm. 184 (2013) 1605; Comput. Phys. Commun. 212 (2017) 239. 

\bibitem{HDECAY}  A.~Djouadi, J.~Kalinowski and M.~Spira, Comput. Phys. Commun. 108 (1998) 56; A. Djouadi, M. Muhlleitner and M. Spira, Acta. Phys. Polon. B38 (2007) 635; A.~Djouadi, J.~Kalinowski, M. Muhlleitner and M.~Spira Comput. Phys. Commun. 238 (2019) 214-231. 

\bibitem{Planck:2018vyg} N.~Aghanim et al.  [Planck Collaboration], Astron. Astrophys. 641 (2020), A6
[erratum: Astron. Astrophys. \textbf{652} (2021), C4]. 

\bibitem{MicroMegas} G.~Belanger, F.~Boudjema, A.~Pukhov and A.~Semenov, Comput. Phys. Commun. 176 (2007) 367.

\bibitem{DD-xsection} G. Belanger, F. Boudjema, A. Pukhov and A. Semenov, Comput. Phys. Commun. 180 (2009) 747.

\bibitem{Manconi:2025ogr} S.~Manconi, C.~Eckner, F.~Calore and F.~Donato,  arXiv:2511.03350 [hep-ph].


\bibitem{LHC-H-WG} D.~de Florian \textit{et al.} [LHC Higgs Cross Section Working Group],
CERN Yellow Rep. Monogr. 2 (2017), 1-869.

\bibitem{CMS:2022goy} A.~Tumasyan et al. [CMS collaboration], JHEP {07} (2023) 073.

\bibitem{Anatomy1} A.~Djouadi,  Phys. Rept. 459 (2008) 1. 

\bibitem{Choi:2021nql}
S.Y.~Choi, J.S.~Lee and J.~Park,
Prog. Part. Nucl. Phys. 120 (2021) 103880. 

\bibitem{ATLAS:2022fpx} G.~Aad et al.,  ATLAS collaboration, Eur. Phys. J. C83 (2023) no.6, 519. 

\bibitem{Benbrik:2025hol}
R.~Benbrik, M.~Boukidi, K.~Kahime, S.~Moretti, L.~Rahili and B.~Taki,
Phys. Lett. B868 (2025) 139688. 


\end{thebibliography}
\end{document}